\documentclass{JHEP3}  
\usepackage{epsfig}
\def\journal#1&#2(#3){\unskip, \sl #1\ \bf #2 \rm(19#3) }
\def\andjournal#1&#2(#3){\sl #1~\bf #2 \rm (19#3) }

\def\eg{{\it e.g.}}

\def\etc{{\it etc}}
\def\p{\partial}

\def\frac#1#2{{#1\over#2}}

\def\inbar{\,\vrule height1.5ex width.4pt depth0pt}
\def\IC{\relax\hbox{$\inbar\kern-.3em{\rm C}$}}
\def\IR{\relax{\rm I\kern-.18em R}}
\def\IP{\relax{\rm I\kern-.18em P}}

\catcode`\@=11
\def\slash#1{\mathord{\mathpalette\c@ncel{#1}}}
\def\OO{{\cal O}}

\def\SS{{\cal S}}

\def\VV{{\cal V}}

\def\underrel#1\over#2{\mathrel{\mathop{\kern\z@#1}\limits_{#2}}}

\catcode`\@=12

\def\exp{{\rm exp}}

\def\[{[}
\def\]{]}

\def\comment#1{ }

\def\draftnote#1{\ifdraft{\baselineskip2ex
                 \vbox{\kern1em\hrule\hbox{\vrule\kern1em\vbox{\kern1ex
                 \noindent \underbar{NOTE}: #1
             \vskip1ex}\kern1em\vrule}\hrule}}\fi}
\def\internote#1{\ifinter{\baselineskip2ex
                 \vbox{\kern1em\hrule\hbox{\vrule\kern1em\vbox{\kern1ex
                 \noindent \underbar{Internal Note}: #1
             \vskip1ex}\kern1em\vrule}\hrule}}\fi}

\def\inbar{\hskip.3em\vrule height1.5ex width.4pt depth0pt}
\def\IC{\relax{\inbar\kern-.3em{\rm C}}}
\def\IN{\relax{\rm I\kern-.16em N}}
\def\IQ{\relax\hbox{$\inbar$\kern-.3em{\rm Q}}}
\def\IZ{\relax{\rm Z\kern-.8em Z}}
\def\be{\begin{equation}}
\def\ee{\end{equation}}
\def\bea{\begin{eqnarray}}
\def\eea{\end{eqnarray}}

\title{Exact potential and scattering amplitudes from the tachyon non-linear $\beta$-function}

\author{Erasmo Coletti\\
{Center for Theoretical Physics}, {Cambridge, MA 02139, U.S.A.} 
\email{E-mail:colettie@mit.edu}}

\author{Valentina Forini\\Dipartimento di Fisica and I.N.F.N. Gruppo
Collegato di Trento, 
Universit\`a di Trento, 38050 Povo (Trento). Italia.
\email{E-mail:forini@science.unitn.it}}

\author{Gianluca Grignani\\Dipartimento di Fisica and Sezione I.N.F.N., 
Universit\`a di Perugia, Via A. Pascoli I-06123, Perugia, Italia.
\email{E-mail:gianluca.grignani$@$pg.infn.it}}
\author{Giuseppe Nardelli\\Dipartimento di Fisica and I.N.F.N. Gruppo
Collegato di Trento, 
Universit\`a di Trento, 38050 Povo (Trento). Italia.
\email{E-mail:nardelli@science.unitn.it}}
\author{Marta Orselli\\Dipartimento di Fisica and Sezione I.N.F.N., 
Universit\`a di Perugia, Via A. Pascoli I-06123, Perugia, Italia.
\email{E-mail:marta.orselli$@$pg.infn.it}}

\abstract{
We compute, on the disk, the non-linear tachyon $\beta$-function, $\beta^T$,
of the open bosonic string theory.
$\beta^T$ is determined both in an expansion to the third power of the field and
to all orders in derivatives and in an expansion
to any power of the tachyon field in the leading order in derivatives.
 We construct the Witten-Shatashvili (WS)
space-time effective action $S$ and prove that it
has a very simple universal form in terms of the renormalized tachyon field and $\beta^T$.
The expression for $S$ is well suited to studying both processes that are
far off-shell, such as tachyon condensation, and close to the mass-shell,
such as perturbative on-shell amplitudes.
We evaluate $S$ in a small derivative expansion,
providing the exact tachyon potential. 
The normalization of $S$ is fixed by requiring that the field
redefinition that maps $S$ into the tachyon effective action derived from the
cubic string field theory is regular on-shell. The normalization factor 
is in precise agreement with the one required for verifying all 
the conjectures on tachyon condensation.
The coordinates in the space of couplings
in which the tachyon $\beta$-function is non linear are the most appropriate to study
RG fixed points that can be interpreted as solitons of $S$, $i.e.$ D-branes.}

\keywords{Tachyon condensation, string field theory}

\begin{document} 

\section{Introduction}

One of the most interesting problems in string theory 
is to understand how the background
space-time on which the string propagates arises in a self-consistent
way.  For open strings, there are two main approaches to this problem,
cubic string field theory  \cite{Witten:1985cc} and background
independent string field theory \cite{Witten:1992qy,Witten:1992cr,
Shatashvili:1993kk,Shatashvili:1993ps}.

Background independent open string field theory has been useful 
for finding the classical tachyon potential energy functional and the
leading derivative terms in the tachyon effective 
action~\cite{Gerasimov:2000zp,Kutasov:2000qp,Tseytlin:2000mt,Grignani:2002rx}. 
It is formulated as a problem in
 boundary conformal
field theory. One begins with the partition function of open-string
theory where the world-sheet is a disk.  The strings in the bulk are
considered to be on-shell and a boundary interaction with arbitrary
operators is added.  The configuration space of open string field
theory is then taken to be the space of all possible boundary
operators modulo gauge transformations  and  field
redefinitions. Renormalization fixed points, which correspond to
conformal field theories, are solutions of classical equations of
motion and should be viewed as the solutions of classical string field
theory.

Due to the existence of a tachyon in the bosonic string theory, 
the 26-dimensional Minkowski space background about which the string
is quantized is unstable. An unstable state should decay to something
and the nature of both the decay process and the endpoint of the decay
are crucial questions.  Some understanding of this
process has been achieved for the open bosonic string.  The key idea
is that of Sen~\cite{Sen:1999mh}.   The open bosonic string tachyon
reflects the instability of the D-25 brane.  This unstable D-brane
should decay by condensation of the open string tachyon field.  The
energy per unit volume released in the decay should be the D-25 brane
tension and the end-point of the decay is the closed string vacuum
\cite{Sen:1999mh,Elitzur:1998va,Harvey:1999gq,Kutasov:2000qp}. There
are also intermediate unstable states which are the D-branes of all
dimensions between zero and 25.

The decay of unstable systems 
of D-branes, pictured as a tachyon field rolling down a potential 
toward a stable minimum, can also be addressed in the context of the
boundary string field theory. It involves deforming the world sheet 
conformal field theory of the unstable D-brane by an exact marginal, 
time dependent tachyon profile. The quantitative study of rolling tachyons 
was initiated by A. Sen
\cite{Sen:2002nu,Sen:2002in,Sen:2002an,Sen:2002vv,Sen:2002qa} and
has been recently used to obtain other forms of tachyon effective 
actions~\cite{
Garousi:2003pv,Garousi:2003ur,Kutasov:2003er,
Fotopoulos:2003yt,Niarchos:2004rw}.
The construction of the space-time tachyon effective 
action in background independent open string field theory
is the subject of the present paper.

By classical power-counting the tachyon field $T(X)$ has dimension one and is
a relevant operator. If $T(X)$ is the only interaction, the field theory
is perturbatively super-renormalizable. 
If $T(X)$ and the other fields are adjusted so that the sigma model
that they define is at an infrared fixed point of the renormalization
group (RG), these background fields are a solution of the classical
equations of motion of string theory.   Witten and
Shatashvili~\cite{Witten:1992qy,Shatashvili:1993kk} have argued that
these equations of motion come from an action which 
can be derived from the disk partition function $Z$ by a
prescription which we shall make use of below.
According to this prescription the effective action for a generic 
coupling constant $g^i$
(which can be identified with the tachyon,
the gauge or any other field that correspond to excitations of the open 
bosonic string)
is related to the renormalized
partition function of open string theory on the disk, $Z(g^i)$, through
\be
S=\left(1-\beta^i\frac{\delta}{\delta g^i}\right)Z(g^i)\ \ ,
\label{S}
\ee
where $\beta^i$ is the beta-function\footnote{In this paper the $\beta$ function
is positive for relevant perturbations. In some other papers on
the subject, \eg \cite{Kutasov:2000qp}, the opposite conventions are used.} of the coupling $g^i$.
Note that (\ref{S}) fixes the additive ambiguity in $S$ by requiring that
at RG fixed points $g^*$, in which $\beta^i(g^*)=0$,
\be
S(g^*)=Z(g^*)\ \ .
\label{fixedpt}
\ee
The derivative of the action $S$ with respect to the coupling constant
$g^i$ must be related to the $\beta$-function through a metric according to
\be
\frac{\partial S}{\partial g^i}=-\beta^j G_{ij}(g)\ .
\label{metric}
\ee
 $G_{ij}$ should be a non-degenerate metric, otherwise there would be an extra 
zero which could not be interpreted as a conformal field theory on the 
world sheet. Eq.(\ref{metric}) indicates that the RG flow is 
actually a gradient flow. The prescription (\ref{S}) provides a definition
of the metric $G_{ij}$ in the space of couplings.

The $\beta$-functions appearing in (\ref{S}) are in general non-linear 
functions of the couplings $g^i$.
 When the linear parts of the $\beta^i$ ({\it i.e}
the anomalous dimensions $\lambda_i$ of the corresponding coupling)
satisfy a so called ``resonant condition'', the non linear parts 
of the $\beta$-function cannot be removed by a coordinate redefinition 
in the space of couplings~\cite{Shatashvili:1993ps}.
 Such resonant condition is nothing but 
the mass-shell condition so that, near the mass-shell, 
the $\beta$-functions are necessarily non-linear.

However, when the resonant condition does not hold,
a possible choice of coordinates on the space
of string fields is one in which the $\beta$-functions are exactly
linear. This choice can always be made locally~\cite{Kutasov:2000qp}
and is well suited to studying processes which are far
off-shell, such as tachyon condensation.
These coordinates, however, 
become singular when the components of the string field
($e.g.$ $T(X)$, $A_\mu(X)$ \etc.) go on-shell.  These coordinates can be used
to construct, for example, the tachyon effective potential, but become
singular when one tries to derive
an effective action which reproduces the on-shell amplitudes.
In particular, if the Veneziano amplitude
needs to emerge from the tachyon effective action it is necessary to consider the 
whole non-linear $\beta$-function in (\ref{S}).
A complete renormalization of the theory in fact
makes the $\beta$-function non-linear in $T(X)$~\cite{Klebanov:1987wx} 
so that, since the vanishing of the
$\beta$-function is the field equation for $T$, 
these nonlinear terms describe tachyon scattering. 
One of the goal of this paper is to construct non-linear expressions for 
the $\beta$-functions which are valid away from the RG fixed point.
With these expressions for the non-linear tachyon $\beta$-function we shall construct the
Witten-Shatashvili (WS) space-time action (\ref{S}).
We shall prove that (\ref{S})
has the following very simple form in the coupling space coordinates in which the tachyon
$\beta$-function is non-linear
\be
S=K\int d^{26}X\left[1-T_R(X)+\beta^T(X)\right]\ \ ,
\label{sws}
\ee
where $T_R$ is the renormalized tachyon field and $K$ is a constant related to the
D25-brane tension. This formula is universal as it
does not depend on how many couplings are switched on. Eq. (\ref{sws}) arises from
the expression that links the renormalized tachyon field to the partition function
that appears in (\ref{S}), namely 
 $Z=K\int d^{26}X(1-T_R)$. $T_R$ is then a non-linear function of 
the bare coupling $T$ and in these coordinates the $\beta$-function is non-linear.
When couplings other than the tachyon 
are introduced in $Z$,  $\beta^T$ will depend on them so that $S$ will
provide the space-time effective action also for these couplings.

With this prescription we shall compute the non-linear $\beta$-function $\beta^T$ for the tachyon 
field up to the third order 
in powers of the field and to any order in derivatives of the field.
 From this we shall show that the solutions of the RG fixed point equations
generate the three and four-point open bosonic string scattering amplitudes involving 
tachyons. Then, with the same renormalization prescription, we shall compute $\beta^T$
 to the leading orders in derivatives but to
any power of the tachyon field  and we shall show that $S$ 
coincides with the one-found in~\cite{Gerasimov:2000zp,Kutasov:2000qp,Tseytlin:2000mt}. 
Obviously, $S$ up to the first three powers of $T$ and expanded to the leading order 
in powers of derivatives can be obtained from both calculations and the results coincide.

In the case of profiles $T_R(k)$ that have support near the on-shell momentum $k^2\simeq 1$
the equation $\beta^T(k)=0$ can be derived as the equation of motion of an action.
We shall show that this action coincides with the tachyon effective action computed, for the 
almost on-shell profiles, form the cubic string field theory up to the fourth power 
of the tachyon field.

The knowledge of the non-linear tachyon $\beta$-function is very important also for another reason.
The solutions of the equation $\beta^T=0$ give the conformal fixed points, the backgrounds that
are consistent with the string dynamics. In the case of slowly varying tachyon profiles,
we shall show that the equations of motion for the WS action can be made
identical to the RG fixed point equation  $\beta^T=0$. We shall find solutions of this equation
to which correspond a finite value of the WS action.
Being solutions of the RG equations, these solitons are lower dimensional D-branes for which the
finite value of $S$ provides a quite accurate prediction of the D-brane tension. 

We shall also show that
the WS action constructed in terms of a linear $\beta$-function~\cite{Frolov:2001nb}
is related to the action  (\ref{sws}) by a field redefinition, 
and that this field redefinition becomes singular on-shell. This 
is in agreement with the 
Poincar\'e-Dulac theorem~\cite{Zamolodchikov:1989zs} 
used in~\cite{Shatashvili:1993ps} to prove that when the resonant condition 
holds, namely near the on-shellness, the $\beta$-function has to be non-linear.

The tachyon effective action up to the third power
in the fields is known exactly also from the cubic string 
field theory~\cite{Witten:1985cc}. 
This raises the interesting question of how 
the action $S$ obtained in this paper
is related to the cubic SFT.
It seems clear that the cubic SFT must correspond to (\ref{S},\ref{metric})
for a particular choice of coordinates on the space of string fields (or
worldsheet couplings).  The  two sets of coordinates are related by
a complicated transformation which we shall derive in this paper. 
The cubic SFT parametrization of worldsheet RG is
regular close to the mass shell. It 
very well reproduces tachyon scattering~\cite{Taylor:2002fy},
to it must correspond a non-linear beta-function.
Thus a coordinate transformation that 
relates the two effective actions needs a 
non-linear beta function in the definition (\ref{S}).
We shall show that this field redefinition exists and that
it is non-singular on-shell
only when $K$ in (\ref{sws}) coincides with the tension of the
D25-brane, in agreement
with all the conjectures involving tachyon condensation~\cite{Sen:1999mh,Sen:1999xm,Sen:1998sm}.

\section{Boundary string field theory}

In Witten's construction of open boundary string field theory~\cite{Witten:1992qy}
the space of all two dimensional worldsheet field theories
on the unit disk, which are conformal in the interior of the disk
but have arbitrary boundary interactions, is described by the world-sheet action 
\be
{\SS=\SS_0+\int_0^{2\pi}{d \tau\over2\pi}\VV \label{SS}}
\ee
where $\SS_0$ is a free action describing an open plus closed
conformal background and $\VV$ is a general perturbation defined on the
disk boundary.
We will discuss the twenty six dimensional bosonic string,
for which (\ref{SS}) can be expressed in terms of a
derivative expansion (or level expansion) of the form
\be
{\VV=T(X)+A_\mu(X)\partial_\tau
X^\mu+B_{\mu\nu}(X)\partial_\tau
X^\mu\partial_\tau X^\nu+C_\mu(X)
\partial^2_\tau X^\mu+\cdots}
\label{Nu}
\ee
Without the perturbation $\VV$
the boundary conditions on $X$
are $\p_r X^\mu\vert_{r=1}=0$, where $r$ is the radial variable
on the disk.

$\VV$ is a ghost number zero operator and it is useful to introduce a ghost number
one operator $\OO$ via
\be
{\VV=b_{-1}\OO.}
\ee
We shall consider the simplest case in which ghosts decouple from matter
so that, as in (\ref{Nu}), $\VV$ is constructed out of matter fields alone 
\be
\OO=c\VV \ .
\label{OO}
\ee
The space-time string field theory
action $S$ is defined through its derivative $dS$ 
which is a two point
function computed with the worldsheet action (\ref{SS}).
More generally one can introduce some basis elements $\VV_i$
for operators of ghost number $0$ so that
the space of boundary perturbations $\VV$ can be parametrized as
\be{\VV=\sum_i g^i \VV_i}
\ee
where the coefficients $g^i$ are couplings on the world-sheet theory,
which are regarded as fields from the space-time point of view, and
$\OO=\sum_i g^i \OO_i$.
In this parametrization the space-time action is defined through its 
derivatives with respect 
to the couplings and has the form
\be{ {\partial S\over \partial g^i}={K\over2}
\int_0^{2\pi}{d\tau\over2\pi}\int_0^{2\pi}{d\tau^\prime\over2\pi}
\langle\OO_i(\tau)\{Q,\OO(\tau^\prime)\}\rangle_g}\ \ ,
\label{sft}
\ee
where $Q$ is the BRST charge and the correlator is evaluated
with the full perturbed
worldsheet action $\SS$. 

If $\VV_i$ is a conformal primary field of dimension $\Delta_i$,
for $\OO$'s of the form (\ref{OO}), one has
\be
{\{Q, c\VV_i\}=(1-\Delta_i)c\partial_\tau c\VV_i}\ ,
\ee
so that from (\ref{sft}) one gets
\be
{{\partial S\over \partial g^i}=-(1-\Delta_j) g^j
G_{ij}(g)} \ ,
\label{dsgi}
\ee
where
\be{G_{ij}=2 K
\int_0^{2\pi}{d\tau\over2\pi}\int_0^{2\pi}{d\tau^\prime\over2\pi}
\sin^2({\tau-\tau^\prime\over 2})\langle
\VV_i(\tau) \VV_j(\tau^\prime)
\rangle_g}\ .
\label{gij}
\ee
Eq.(\ref{dsgi}) cannot be true in general,
since it does not transform covariantly under reparametrizations
of the space of theories, $g^j\to f^j(g^i)$. Indeed, 
$\partial_i S$ and $G_{ij}$ transform as tensors,
(the latter is the metric on the space of worldsheet
theories), but $g^i$ does not.

The correct   covariant generalization of (\ref{dsgi}) was
given in~\cite{Shatashvili:1993kk,Shatashvili:1993ps}.
The worldsheet RG defines a natural vector field on the space of
theories: the $\beta$-function $\beta^i(g)$,
which transforms as a covariant vector under
reparametrizations of $g^i$.
The covariant form of (\ref{dsgi}) is thus (\ref{metric}).
If we assume that total derivatives inside the correlation function
decouple and that there are no contact terms,  it turns out that 
the $\beta$-function in (\ref{S}) is the linear $\beta$-function.
This implies that
the equations of motion derived from the action (\ref{S}) are just linear.
However, as shown by Shatashvili~\cite{Shatashvili:1993kk,Shatashvili:1993ps},
contact terms show up in the computation on the world-sheet and cannot 
be ignored.
The point is that the operator $Q$, which is constructed out of the BRST
operator in the bulk and should be independent on the couplings because
the perturbation is on the boundary, actually depends on the couplings when the
contour integral approaches the boundary of the disk. 
A way to fix the structure 
of the contact terms is to consider that, since $dS$ is a one-form,
the derivative of $dS$ should be zero independently of the choice of 
the contact terms that one makes in the computation. 
This leads to the following formula for the
vector field in equation (\ref{S}) 
\be
\beta^i=(1-\Delta_i)g^i+\alpha^i_{jk}g^jg^k+\gamma^i_{jkl}g^jg^kg^l+
\cdots
\label{nlb}
\ee
This is an expression for the $\beta$-function with all the non-linear terms.
According to the Poincare-Dulac Theorem about vector fields 
(whose relevance to the $\beta$-function related issues was stressed many 
times by Zamolodchikov~\cite{Zamolodchikov:1989zs})
 every vector field can be linearized
by an appropriate redefinition of the coordinates up to the resonant term.
In the second order of equation (\ref{nlb}) the resonance condition 
is given by
\be
\Delta_j+\Delta_k-\Delta_i=1\ .
\label{resonant}
\ee
The resonance condition means that the $\beta$-function cannot be linearized
by a coordinate transformation and that all the non-linear 
terms cannot be removed from the $\beta$-function equation (\ref{nlb}).
When $g^i$ is the tachyon field $T(k)$,
the resonant condition (\ref{resonant}) corresponds to the mass-shell conditions for three 
tachyons.
We shall prove in what follows that the WS action $S$ up to the third order 
in the tachyon fields, constructed in terms of the linear 
$\beta$-function~\cite{Frolov:2001nb}, is related to the $S$
made of a non-linear $\beta$-function by a field redefinition, but that
this field redefinition becomes singular on-shell.

\section{Integration over the bulk variables}

Let us now restrict ourselves to the specific example of 
open strings propagating in a tachyon background. The partition function reads
\begin{equation}
Z=\int [d X^\mu(\sigma,\tau)]\exp\left( -S[X]\right)\ ,
\label{partf}
\end{equation}
where the action is
\begin{equation}
S[X]=\int d\sigma d\tau \frac{1}{4\pi}\partial_a X(\sigma,\tau)
\cdot \partial_a X(\sigma,\tau)
+\int_0^{2\pi}\frac{d\tau}{2\pi} T(X(\tau))\ .
\label{actionint}
\end{equation}
Here, the first term in (\ref{actionint}) is the bulk action and is 
integrated over the
volume of the unit disk.  The second term in (\ref{actionint})
is integrated on the circle which is the boundary of the unit disk and
describes the interactions.
The scalar fields $X^\mu$ have $D$ components with $\mu=1,...,D$ and
we shall assume 
$D=26$ in what follows for a critical string. We are working in a system of units 
where $\alpha'=1$.

We begin with the observation that the
bulk excitations can be integrated out of (\ref{partf}) to get an
effective non-local field theory which lives on the boundary~\cite{Callan:nz}.  To do
this we write the field in the bulk as~\cite{Grignani:2002rx}
$$
X=X_{\rm cl}+X_{\rm qu}\ ,
$$
where $$\partial^2X_{\rm cl}=0$$ and $X_{\rm cl}$ approaches the
fixed (for now) boundary value of $X$,
$$X_{\rm cl}\to X_{\rm bdry}~{\rm and}~X_{\rm qu}\to 0\ .$$ 
Then, in the
bulk, the functional measure is $dX=dX_{\rm qu}$ and
\begin{equation}
S=\int \frac{d^2\sigma}{4\pi}\partial X_{\rm qu}\cdot\partial X_{\rm
qu} + \int \frac{d\tau}{2\pi} \left\{ \frac{1}{2}X^\mu |i\partial_\tau|X^\mu
+T(X) \right\}\ ,
\label{1daction}
\end{equation}
where we omitted the {\rm cl} index in the last integral.
Then, the integration of $X_{\rm qu}$ produces a multiplicative
constant in the partition function - the partition function of the
Dirichlet string, which we shall denote $K$.  The kinetic term
in the boundary action is non-local.  The absolute value of the
derivative operator is defined by the Fourier transform,
$$
|i\partial_\tau|\delta(\tau-\tau')=\sum_n\frac{|n|}{2\pi}e^{in(\tau-\tau')}\ .
$$
The partition function of the boundary theory is then
\begin{equation}
Z(J)=K\int [dX_\mu]e^{-\int_0^{2\pi}\frac{d\tau}{2\pi}
\left(\frac{1}{2}X^\mu |i\partial| X^\mu + T(X)-J\cdot X\right)}\ ,
\label{zj}
\end{equation}
where we have added a source $J^\mu(\tau)$ so that the path integral can
be used as a generating functional for correlators of the fields $X^\mu$
restricted to the boundary.  In particular, this source will allow us
to compute the correlation functions of vertex operators of open
string degrees of freedom.  The remaining path integral over the
boundary $X^\mu(\tau)$ defines a one-dimensional field theory with
non-local kinetic term.  If the tachyon field were absent ($T=0$), the
further integration over $X^\mu(\tau)$ would give a factor which
converts the Dirichlet string partition function to the Neumann string
partition function.

\section{Partition function on the disk and the renormalized tachyon field}

When only the tachyon field is considered as a boundary perturbation,
the Witten-Shatashvili action is given by
\begin{equation}
S=\left( 1-\int \beta^T\frac{\delta}{\delta T}\right)Z\ ,
\label{witten}
\end{equation}
where $Z$ is the partition function of the boundary theory on the disk 
and $\beta^T$ is the tachyon $\beta$-function.
It is useful to introduce a constant source term $k$ for the 
zero mode of the $X$ field, the integral over the zero mode 
variable will just provide the
energy-momentum conservation $\delta$-function.
The partition function (\ref{zj}) in the presence of this constant source reads
\begin{equation}
Z(k)=K\int [dX_\mu]e^{-\int_0^{2\pi}\frac{d\tau}{2\pi}
\left(\frac{1}{2}X^\mu |i\partial_\tau| X^\mu + T(X)-ik\cdot \hat X\right)}\ ,
\label{zetak}
\end{equation}
where $\hat X$ is the zero mode which is defined by
\be
\hat X^\mu = \int_0^{2\pi}\frac{d\tau}{2\pi} X^\mu(\tau)\ .
\label{zerom}
\ee
In this section  we shall expand the exponential in eq.(\ref{zetak})
in powers of $T(X)$.
The first non-trivial term is
\be
Z^{(1)}(k)=-K\int [dX_\mu]\int dk_1
\int_0^{2\pi}\frac{d\tau_1}{2\pi}T(k_1)e^{-\int_0^{2\pi}\frac{d\tau}{2\pi}
\left(\frac{1}{2}X^\mu  |i\partial_\tau| X^\mu\right)-i k \hat X+ik_1 X(\tau_1)}\ .
\ee
The functional integral over the non-zero modes of $X(\tau)$ gives
\be
{Z^{(1)}(k)}=-K\int d\hat X_\mu\int dk_1
T(k_1)e^{-\frac{k_1^2}{2} G(0)
+i (k_1-k)\hat X}\ ,
\label{z1}
\ee
where $G(\tau)$ is the Green function of the operator $|i\partial_\tau|$
\be
G(\tau_1 -\tau_2)=2\sum_{n=1}^{\infty}e^{\epsilon n}
\frac{\cos n\left(\tau_1 -\tau_2\right)}{n}=
-\log \left[1-2e^{-\epsilon}\cos \left(\tau_1 -\tau_2\right)
+e^{-2\epsilon}\right]
\label{prop}
\ee
and $\epsilon$ is an ultraviolet cut-off.
In all the calculations we shall use the following prescription for $G(\tau)$
\bea
G(\tau)=\cases{
-\log\left[ c\sin^2\left(\frac{\tau}{2}\right)\right] & $\tau\ne 0$ \cr
-2\log\epsilon &$\tau=0$}\ .
\label{prop0}
\eea 
The coefficient $c$ reflects the ambiguity involved in subtracting
the divergent terms. Its value is scheme dependent and should be fixed 
by some renormalization prescription. We choose the value $c=4$ for 
later convenience.
This arbitrariness was discussed
in~\cite{Tseytlin:2000mt, Grignani:2002rx}.
The integrals over the zero-modes in eq.(\ref{z1}) 
give a $26$-dimensional $\delta$-function so that 
\be
-Z^{(1)}(k)=K T(k)
\epsilon^{k^2 -1}
\ee
and we can identify
\be
T_R(k)\equiv T(k)\epsilon^{k^2 -1}=-\frac{Z^{(1)}(k)}{K}\ .
\label{tr1}
\ee
This equation provides the renormalized coupling $T_R$ 
in terms of the bare coupling $T$ to the 
lowest order in perturbation theory.
$1-k^2$ is the anomalous dimension of the tachyon field.
The second order term in $T$ is given by
\be
Z^{(2)}(k)=K
\int_0^{2\pi}
\frac{d\tau_1}{4\pi}
\frac{d\tau_2}{2\pi}
\int dk_1dk_2 {T}(k_1){T}(k_2)
\left<e^{ik_1X(\tau_1)}e^{ik_2X(\tau_2)}e^{-i k\hat{X}}\right> \ .
\label{z2}
\ee
Again in (\ref{z2}) the integral over the zero modes $\hat X^\mu$ gives just a
$26$-dimensional $\delta$-function, $\delta \left(k-k_1-k_2\right)$,
and we can perform the integral over the non-zero modes of $X(\tau)$ to get
\bea
Z^{(2)}(k)=&& K
\int_0^{2\pi}
\frac{d\tau_1}{4\pi}
\frac{d\tau_2}{2\pi}
\int dk_1dk_2(2\pi)^{D}\delta \left(k-k_1-k_2\right) 
{T}(k_1){T}(k_2)\cr
&&\exp \left[-\frac{1}{2}\left(k_1^2 +k_2^2\right)G(0)
-k_1k_2G\left(\tau_1 -\tau_2 \right)\right] \ .
\label{ze2}
\eea
The integral in (\ref{ze2}) becomes
\bea
Z^{(2)}(k)=&&
K
\int dk_1dk_2(2\pi)^{D}\delta \left(k-k_1-k_2\right) 
\epsilon^{k_1^2+k_2^2-2}
T(k_1)T(k_2)\cr
&&\int_0^{2\pi}\frac{d\tau_1}{4\pi}
\frac{d\tau_2}{2\pi}\left[4\sin^2\left(\frac{\tau_1 -\tau_2}{2}\right)
\right]^{k_1k_2}\ .
\label{x}
\eea
The integral over the relative variable $x=(\tau_1-\tau_2)/{2}$ does not
need regularization, it converges when $1+2k_1k_2>0$, providing the result
\be
Z^{(2)}(k)=
\frac{K}{2}
\int dk_1dk_2(2\pi)^{D}\delta \left(k-k_1-k_2\right) \epsilon^{k_1^2+k_2^2-2}
{T}(k_1){T}(k_2)\frac{\Gamma \left(1+2k_1k_2\right)}
{\Gamma^2 \left(1+k_1k_2\right)}\ .
\label{secord}
\ee
The integrand in (\ref{secord}) can be analytically continued also
to the region where $1+2k_1k_2<0$, so that the integral can be performed.
 
To the second order in perturbation theory the renormalized
coupling in terms of the bare coupling reads
\bea
&&T_R(k)=-\frac{Z^{(1)}(k)+Z^{(2)}(k)}{K}\cr
&&=\epsilon^{k^2 -1}\left[T(k)-\frac{1}{2}
\int dk_1dk_2(2\pi)^{D}\delta \left(k-k_1-k_2\right) 
{T}(k_1){T}(k_2)\epsilon^{-(1+2k_1k_2)}\frac{\Gamma \left(1+2k_1k_2\right)}
{\Gamma^2 \left(1+k_1k_2\right)}\right] \ .\cr
&&
\label{tr2}
\eea
The third order contribution to the partition function is given by
\be
{Z^{(3)}(k)}=
-\frac{K}{3!}
\int dk_1dk_2dk_3(2\pi)^{D}\delta \left(k-\sum_{i=1}^3k_i\right)\epsilon^{\sum_{i=1}^3k_i^2-3} 
T(k_1)T(k_2)T(k_3)I(k_1,k_2,k_3)\ ,
\label{z3}
\ee
where $I(k_1,k_2,k_3)$ is the integral
\bea
I(k_1,k_2,k_3)=&&
\frac{2^{2k_1k_2+2k_2k_3+2k_1k_3}}{(2\pi)^3}
\int_0^{2\pi}d\tau_1d\tau_2d\tau_3\
\left[\sin^2\left(\frac{\tau_1 -\tau_2}{2}\right)
\right]^{k_1k_2}\cr
&&\left[\sin^2\left(\frac{\tau_2 -\tau_3}{2}\right)
\right]^{k_2k_3}\left[\sin^2\left(\frac{\tau_1 -\tau_3}{2}\right)
\right]^{k_1k_3}\ .
\label{I1}
\eea
The complete computation of $I(k_1,k_2,k_3)$ will
be given in Appendix A.
The result is given by the completely symmetric formula
\be
I(a_1,a_2,a_3)=
\frac{\Gamma(1+a_1+a_2+a_3)\Gamma(1+2a_1)\Gamma(1+2a_2)\Gamma(1+2a_3)}
{\Gamma(1+a_1)\Gamma(1+a_2)\Gamma(1+a_3)\Gamma(1+a_1+a_2)\Gamma(1+a_2+a_3)
\Gamma(1+a_1+a_3)}\ ,
\label{i}
\ee
where we have set $a_1=k_1k_2$, $a_2=k_2k_3$ and $a_3=k_1k_3$. 
The integral (\ref{I1}) converges when
$1+a_1+a_2+a_3>0$, but its result (\ref{i}) can be analytically continued also 
outside this convergence region. The result (\ref{i}) is in agreement
with the one obtained, with a different procedure, in~\cite{Frolov:2001nb}
but does not coincide with the one provided in the appendix of ref.~\cite{Kutasov:2000qp}.
 Up to the third order in powers of $T$ and to all orders in $k_i$
the relation between the bare and the
renormalized couplings reads
\bea
&&T_R(k)=-\frac{Z^{(1)}(k)+Z^{(2)}(k)+Z^{(3)}(k)}{K}\cr
&&=\epsilon^{k^2 -1}\left[T(k)-\frac{1}{2}
\int dk_1dk_2(2\pi)^{D}\delta \left(k-\sum_{i=1}^3k_i\right) 
{T}(k_1){T}(k_2)\epsilon^{-(1+2k_1k_2)}\frac{\Gamma \left(1+2k_1k_2\right)}
{\Gamma^2 \left(1+k_1k_2\right)} \right. \cr
&&\left. +
\int dk_1dk_2dk_3\frac{(2\pi)^{D}}{3!}\delta \left(k-\sum_{i=1}^3k_i\right) 
{T}(k_1){T}(k_2)T(k_3)\epsilon^{-2(1+\sum_{i<j}k_ik_j)}
I(k_1,k_2,k_3)\right]\ .\cr
&&
\label{tr3}
\eea
In section 6 we shall use this expression to construct the non-linear $\beta$-function.

The renormalized tachyon field can be constructed to all powers 
of the bare tachyon field in the case in which the tachyon profile 
appearing in (\ref{zetak}) is a slowly varying function of $X^\mu$.
In this case one can consider an expansion of (\ref{zetak}) in powers
of derivatives of $T$. To this purpose consider the $n$-th term in the
expansion of (\ref{zetak}) in powers of $T(X(\tau))$, $Z^{(n)}(k)$. Taking 
the Fourier transform of the tachyon field and performing all the contractions
of the $X(\tau_i)$ fields, for $Z^{(n)}(k)$ we get
\bea
&&Z^{(n)}(k)=K\frac{(-1)^n}{n!}\epsilon^{-n}\int 
\prod_{i=1}^{n}dk_i T(k_i)\int_0^{2\pi}\prod_{i=1}^{n}\left(\frac{d\tau_i}{2\pi}\right)
\cr
&&e^{-\sum_{i=1}^n\frac{k_i^2}{2}G(0)-\sum_{i<j}^nk_ik_jG\left(\tau_i-\tau_j\right)}
\delta\left(k-\sum_{i=1}^nk_i\right)\ .
\label{zetan}
\eea
Note that with our regularization prescription
the dependence on the cut-off in (\ref{zetan}) comes only from
the zero distance propagator $G(0)$ and from the explicit scale dependence
of the tachyon field.
If the tachyon profile is a slowly varying function of $X^\mu$ we can
expand inside the integrand of (\ref{zetan}) in powers of the momenta $k_i$.
The leading and next to leading terms in this expansion read
\bea
Z^{(n)}(k)=&&K \frac{(-1)^n}{n!}\prod_{i=1}^{n}\int dk_i\delta\left(k-\sum_{i=1}^nk_i\right)
\epsilon^{-n}\prod_{i=1}^n T(k_i)\cr
&&\left(1+\sum_{i=1}^nk_i^2\log\epsilon
+\sum_{i<j}^nk_ik_j\log{\frac{c}{4}}\right)\ ,
\label{zetanexp}
\eea
where the last term comes from the integral over a couple of $\tau$ variables
of the propagator $G\left(\tau_i-\tau_j\right)$,
the other integrations over $\tau_k$ $k\ne i,j$ being trivial.
Here we have kept explicit the ambiguity $c$ appearing in the propagator (\ref{prop0}) 
to show that the result greatly simplifies with the choice $c=4$.
Unless otherwise stated, we shall adopt this choice
throughout the paper.
As before, the renormalized tachyon field $T_R(k)$ can be obtained from (\ref{zetanexp})
by summing over $n$ from 1 to $\infty$, changing sign and dividing by $K$.
Taking the Fourier transform of $T_R(k)$ with $c=4$,
to all orders in the bare tachyon field and to the leading order in derivatives, we get the 
renormalized tachyon field $T_R(X)$
\be
T_R(X)=1-\exp\left\{-\frac{1}{\epsilon}\left[T(X)-
\triangle T(X)\log\epsilon\right]\right\} \,
\label{trsum}
\ee
where $\triangle$ is the Laplacian. Again in section 6 we shall use this expression to compute
the non-linear tachyon $\beta$-function.

{}From eqs.(\ref{tr1},\ref{tr2},\ref{tr3},\ref{trsum}) it is clear that 
the general relation between the renormalized tachyon field $T_R(X)$ 
and the partition function $Z\equiv Z(k=0)$ is simply
\be
Z=K\int d^{26}X\left[1-T_R(X)\right]\ .
\label{zt}
\ee
This expression is true also when other couplings are present. $T_R$
in this case would be a non linear function also of the other bare
couplings but its relation with the partition function of the
theory would always be given by (\ref{zt}). We shall prove eq.(\ref{zt}) in the
next section.

\section{Background-field method}

The partition function of the boundary theory on the disk in general is given by
\begin{equation}
Z=K\int [dX_\mu]e^{-
\left(S_0[X] + \int_0^{2\pi}\frac{d\tau}{2\pi}\VV[X(\tau)]\right)}\ ,
\label{zbfm}
\end{equation}
where $S_0=\int d\tau X^\mu|i\partial_\tau|X^\mu$ and $\VV[X(\tau)]$ is given in (\ref{Nu}).
Our goal is to determine the relationship between the renormalized and the bare couplings
of the one-dimensional field theory. To this purpose we shall make use of the 
background field method~\cite{Klebanov:1987wx}. We expand the fields $X^\mu$ around a classical background
$X_0^\mu$ which satisfies the equations of motion and which varies slowly compared to the cut-off scale,
$$
X^\mu=X_0^\mu+Y^\mu\ .
$$ 
The effective action is $S_{\rm eff}[X_0]=-\log Z[X_0]$
and the aim of the renormalization process is to rewrite
the local terms of $S_{\rm eff}[X_0]$ in terms of renormalized couplings in such a way that
$S_{\rm eff}[X_0]$ has the same form of the original action
\be
 S_{\rm eff}[X_0]\biggr |_{\rm local}=S_0[X_0]+
  \int_0^{2\pi}\frac{d\tau}{2\pi}\VV_R[X_0(\tau)]\ .
\ee
$Z[X_0]$ can be conveniently calculated in powers
of the boundary interaction $\VV$.
The first order for example reads, up to the multiplicative constant $K$,
\bea
-\int_0^{2\pi}\frac{d\tau}{2\pi}\int dk e^{ikX_0}&&\left<\left[T(k)+A_\mu(k)\partial_\tau
(X_0^\mu+Y^\mu)+B_{\mu\nu}(k)\partial_\tau(X_0^\mu+Y^\mu)\partial_\tau(X_0^\nu+Y^\nu)\right.\right.\cr
&&\left.\left.
+C_\mu(k)
\partial^2_\tau(X_0^\mu+Y^\mu) +\cdots\right]e^{ik Y}\right>\ .
\label{1stnu}
\eea
The renormalized couplings $T_R(k)$ will be given
by the opposite of the coefficient of the term in (\ref{1stnu}) 
 that does not contain $X_0$  derivatives.
Analogously, the renormalized $A^R_\mu(k)$ will be determined
by the coefficient of $\partial_\tau X^\mu_0$,  $B^R_{\mu\nu}(k)$ 
by the coefficient of $\partial_\tau X^\mu_0\partial_\tau X^\nu_0$ and so on.
The second order term in the expansion of $Z[X_0]$ is
\bea
&&\int_0^{2\pi}\frac{d\tau_1}{2\pi}\int_0^{2\pi}\frac{d\tau_2}{4\pi}\int dk_1 dk_2
e^{ik_1X_0(\tau_1)+ik_2 X_0(\tau_2)}\left<e^{ik_1 Y(\tau_1)+ik_2 Y(\tau_2)}\right.\cr
&&\left.\right.\cr
&&\left.\left[T(k_1)+A_\mu(k_1)\partial_{\tau_1}
(X_0^\mu+Y^\mu)+\dots\right]\left[T(k_2)+A_\nu(k_2)\partial_{\tau_2}
(X_0^\nu+Y^\nu)+\dots\right]\right>\ .
\label{sozx0}
\eea
An expansion of the background field $X_0$ in powers of its derivatives is required 
to determine the coefficients of 1, $\partial_\tau X^\mu_0$,  
$\partial_\tau X^\mu_0\partial_\tau X^\nu_0$, $\dots$,
\be
X_0(\tau_2)=X_0(\tau_1)+(\tau_2-\tau_1)\partial_{\tau_1} X_0(\tau_1)+\dots\ .
\label{x0exp}
\ee
If we are interested in renormalization of couplings of the form $\exp[i k X_0]$, namely
in the renormalized tachyon field $T_R(k)$, we can disregard the terms in (\ref{x0exp},\ref{sozx0})
involving derivatives acting on $X_0$.
For example, at the second order,
the only non-vanishing terms in $T$ and $A_\mu$ contributing to $T_R$ are
\bea
T_R(k)=-&&\int dk_1\int dk_2\delta(k-k_1-k_2)
\int_0^{2\pi}\frac{d\tau_2}{4\pi}\left<e^{ik_1 Y(\tau_1)+ik_2 Y(\tau_2)}\right.\cr
&&\left. \right. \cr
&&\left.\left[T(k_1)T(k_2)+A_\mu(k_1) A_\nu(k_2)\partial_{\tau _1}Y^\mu\partial_{\tau_2}
Y^\nu+\dots\right]\right>\ ,
\label{tr2aa}
\eea
where the correlator does not depend on $\tau_1$  since the propagator (\ref{prop0})
of $X(\tau)$ and its derivatives are periodic functions on the unit circle. 
It is not difficult to see that $T_R(k)$ in (\ref{tr2aa}) coincides with the opposite of the 
second order term in the expansion of the partition function
\begin{equation}
Z(k)=\int [dY_\mu]e^{-
\left(S_0[Y] + \int_0^{2\pi}\frac{d\tau}{2\pi}\VV[Y(\tau)]\right)-ik\hat Y}
\label{zbfmy}
\end{equation}
in powers of the couplings. Here $k$ is a constant source for the 
zero mode of the $Y^\mu$ field, $\hat Y^\mu$ (\ref{zerom}). Such a constant
source will just provide the $\delta$-function in (\ref{tr2aa}) 
that imposes the energy-momentum conservation.
This will be true at any order in the expansion in powers of the coupling fields.
Therefore, to all orders in whatever coupling,
the expression for the renormalized tachyon field $T_R(X)$ is related to the 
partition function $Z=Z(k=0)$ precisely by (\ref{zt}), which is the relation that 
we wanted to prove. Note that $T_R$ depends not only on the bare tachyon 
field but also on the other coupling fields (in particular 
$T_R$ will exists also if one starts from
a boundary interaction that does not contain the bare tachyon).
As a consequence, the tachyon $\beta$ function will contain for example 
also the gauge field~\cite{Kostelecky:1999mu}, and this is as it should be, 
since the solution of the equation $\beta^T=0$ will then describe 
the scattering of a tachyon by other excitations (\eg\  from (\ref{tr2aa})
by two vector fields).

\section{$\beta$-function}

In this section we shall perform a calculation of the non-linear tachyon $\beta$-function.
The resulting expression  will then be 
used to derive the Witten-Shatashvili action (\ref{sws},\ref{witten}).
Following~\cite{Klebanov:1987wx}, the most general RG equations
for a set of couplings $g^i$ can be written as
\be
\beta^i\equiv\frac{dg^i}{dt} =\lambda_ig^i+\alpha^i_{jk}g^jg^k
+\gamma^i_{jkl}g^jg^kg^l+\cdots\ \ ,
\label{beta}
\ee
where the scale $t$ is $t=-\log \epsilon$, $\lambda_i$
are the anomalous dimensions corresponding to the couplings $g^i$ 
and there is no summation 
in the first term on the right-hand side. 
This equation has the solution
\be
g^i(t)=e^{\lambda_it}g^i(0)+\left[e^{\left(\lambda_j+\lambda_k\right)t}
-e^{\lambda_it}\right]\frac{\alpha^i_{jk}}{\lambda_j+\lambda_k
-\lambda_i}g^j(0)g^k(0)+b^i_{jkl}(t)g^j(0)g^k(0)g^l(0)+\cdots\ \ ,
\label{gi}
\ee
where $g^i(0)$ are the bare couplings and
\bea
&&b^i_{jkl}(t)g^j(0)g^k(0)g^l(0)=
\left[\left(\frac{2\alpha^i_{jm}\alpha^m_{kl}}{\lambda_j
+\lambda_m-\lambda_i}-\gamma^i_{jkl}\right)\frac{e^{\lambda_it}}
{\lambda_j+\lambda_k+\lambda_l-\lambda_i}\right.\cr
&&\left.\right.\cr
&&\left.+
\left(\frac{2\alpha^i_{jm}\alpha^m_{kl}}{\lambda_k
+\lambda_l-\lambda_m}+\gamma^i_{jkl}\right)\frac{e^{\left(\lambda_j
+\lambda_k+\lambda_l\right)t}}
{\lambda_j+\lambda_k+\lambda_l-\lambda_i}\right.\cr
&&\left.-\frac{2\alpha^i_{jm}\alpha^m_{kl}}
{\left(\lambda_j+\lambda_m-\lambda_i\right)
\left(\lambda_k+\lambda_l-\lambda_m\right)}e^{\left(\lambda_j
+\lambda_k\right)t}\right]g^j(0)g^k(0)g^l(0)\ .
\label{b}
\eea
Let us now consider the case of interest for this paper: open strings propagating 
in a tachyon background. In this case the coupling
$g^i$ is the tachyon field $T(k)$. Then $\lambda_i=1-k^2$ and
$\lambda_j=1-k^2_j$.
Comparing the general solution (\ref{gi}) with eq.(\ref{tr3})
derived in the previous section, we will be able to identify the
renormalized tachyon field in terms of the bare field
up to the third order in powers of the field and to all orders in 
its derivatives.
In the second order term of (\ref{tr3}) the coefficient
proportional to 
$e^{\lambda_it}=\epsilon^{1-k^2}$ appearing in (\ref{beta}) is absent. 
This is due to the fact that the convergence condition
for the integral (\ref{x}), $1+2k_1k_2>0$, implies that 
$\lambda_j+\lambda_k>\lambda_i$ so that in the limit 
$t\to \infty$ the dominant 
contribution comes from $e^{\left(\lambda_j+\lambda_k\right)t}$.
{}From similar arguments, the first and the second terms of the
right-hand side of (\ref{b}) are negligible compared to the
second term, due to the convergence conditions for the integral $I(k_1,k_2,k_3)$ 
computed in the previous section.
This is a general feature of our renormalization procedure.
At the $n$-th order in the bare coupling in the expansion (\ref{gi}),
the renormalized coupling will contain only the term
of the form
\be
e^{t\sum_{k=1}^n\lambda_k} \ .
\ee
This is due to the fact that the integrals over 
the $\tau$'s do not need an explicit regulator, rather 
they can be evaluated in a specific region of the $k_i$ variables
and then analytically continued. Therefore
the only dependence on the cut-off does not come from such integrals 
but from the propagators (\ref{prop0}) evaluated at zero distance. 

Comparing our result for the renormalized tachyon field 
(\ref{tr3}) with the general expressions (\ref{gi},\ref{b}),
for the coefficients in the expansion (\ref{beta}) we find
\bea
&&\alpha^i_{jk}=-\frac{1}{2}\frac{\Gamma(2+2k_jk_k)}{\Gamma^2(1+k_jk_k)}
\delta (k-k_j-k_k)\cr
&&\gamma^i_{jkl}=\frac{1}{3!}\int dk_j dk_k dk_l \delta (k-k_j-k_k-k_l)
\left[2(1+k_jk_k+k_jk_l+k_kk_l)I(k_j, k_k, k_l)\right.\cr
&&\left.-\left(\frac{\Gamma(2+2k_jk_k+2k_jk_l)
\Gamma(1+2k_kk_l)}{\Gamma^2(1+k_jk_k+k_jk_l)\Gamma^2(1+k_kk_l)}
+{\rm cycl.}\right)\right]\ ,
\label{coeff}
\eea
where $I(k_j,k_k,k_l)$ is given in equation (\ref{i}).
The perturbative expression for the $\beta$-function  up to the third 
order in the tachyon field obtained using this procedure therefore is
\bea
&&\beta^T(k)=(1-k^2)T_R(k)-\frac{1}{2}\int dk_1 dk_2
(2\pi)^D \delta (k-k_1-k_2)T_R(k_1) T_R(k_2)
\frac{\Gamma(2+2k_1k_2)}{\Gamma^2(1+k_1k_2)}\cr
&&+\frac{1}{3!}\int dk_1 dk_2 dk_3(2\pi)^D\delta(k-k_1-k_2-k_3)
T_R(k_1) T_R(k_2) T_R(k_3)\cr
&&\left[2(1+k_1k_2+k_1k_3+k_2k_3)I(k_1,k_2,k_3)-\left(\frac{\Gamma(2+2k_1k_2+2k_1k_3)
\Gamma(1+2k_2k_3)}{\Gamma^2(1+k_1k_2+k_1k_3)\Gamma^2(1+k_2k_3)}
+{\rm cycl.}\right)\right].\cr
&&
\label{betafunction}
\eea
We have thus succeeded in deriving a $\beta$-function for tachyon backgrounds which
do not satisfy the linearized on-shell condition.
Exactly the same result can be obtained by taking the derivative of (\ref{tr3})
(or of the opposite of $Z(k)$)
with respect to the logarithm of the cut-off $-\log\epsilon$.
The result obtained in this way must then be expressed in terms of the renormalized field
by inverting (\ref{tr3}) and it coincides with  (\ref{betafunction}).

It is interesting to note that all the known conformal tachyon profiles, like $e^{i X^0}$
or $\cos X^i$ where $i$ is a space index, are solutions of the equation
$\beta^T(X)=0$, where $\beta^T(X)$ is the Fourier transform of (\ref{betafunction}).
These solutions and perturbations around them have been recently used to
construct tachyon effective actions around the on-shellness
~\cite{Garousi:2002wq,Garousi:2003pv, Garousi:2003ur,Kutasov:2003er,Niarchos:2004rw}  
and to study the problem of
the rolling tachyon
~\cite{Sen:2002nu,Sen:2002in,Sen:2002an,Sen:2002vv,Sen:2002qa, Larsen:2002wc, Fotopoulos:2003yt}.

That the non-linear $\beta$-function (\ref{betafunction})
is the correct one can be shown by solving the $\beta^T(k)=0$  equation
perturbatively. The solution of this equation will generate 
the correct scattering amplitudes of open string theory~\cite{Klebanov:1987wx}.
This in turn will show the validity of the general formula (\ref{zt}).
To the lowest order the equation is $(1-k^2)T_0(k)=0$, so that the 
solution $T_0(k)$ satisfies the linearized on-shell 
condition.
By writing $T(k)=T_0(k)+T_1(k)$ and substituting into the equation 
$\beta_T(k)=0$, to the next order we find
\be
T_1(k)=\frac{1}{2}\int dk_1 dk_2
(2\pi)^D\delta(k-k_1-k_2)
T_0(k_1)T_0(k_2)\frac{\Gamma(k^2)}
{\left(1-k^2\right)\Gamma^2\left(k^2/2\right)}\ .
\label{t1}
\ee
The presence of the couplings $T_0$ in (\ref{t1}) sets two of the three
$k_i$ on-shell. To pick out the propagator pole corresponding to the
third $k$ we set it on-shell too.
The scattering amplitude for three on-shell tachyons is given by
the residue of the pole and is $1/2\pi$ with our normalization.

The calculation at the next order proceed in a similar fashion.
One sets $T(k)=T_0(k)+T_1(k)+T_2(k)$ and finds
\bea
&&T_2(k)=-\frac{(2\pi)^D}{3!(1-k^2)}\int dk_1 dk_2 dk_3\delta(k-k_1-k_2-k_3)
T_0(k_1)T_0(k_2)T_0(k_3)I(k_1,k_2,k_3)\cr
&&~~\left\{2\left(1+\sum_{i<j}k_ik_j\right)I(k_1,k_2,k_3)-\left[\frac{\Gamma(2+2k_1k_2+2k_1k_3)
\Gamma(1+2k_2k_3)}{\Gamma^2(1+k_1k_2+k_1k_3)\Gamma^2(1+k_2k_3)}
+{\rm cycl.}\right]\right.\cr
&&~~~\left.-\left[\frac{\Gamma(2+2k_1k_2+2k_1k_3)
\Gamma(2+2k_2k_3)}{\Gamma^2(1+k_1k_2+k_1k_3)\Gamma^2(1+k_2k_3)[1-(k_2+k_3)^2]}
+{\rm cycl.}\right]\right\}\ .
\label{ampl}
\eea
When all the tachyons are on-shell, the last two terms on eq. (\ref{ampl}) cancel and,
as it should be for consistency, the residue of the pole in $k$
 is the scattering amplitude of four on-shell tachyons. 
It is given by 
\be
\frac{\Gamma(1+2k_1k_2)\Gamma(1+2k_2k_3)\Gamma(1+2k_1k_3)}
{\Gamma(1+k_1k_2)\Gamma(1+k_2k_3)\Gamma(1+k_1k_3)\Gamma(1+k_1k_2+k_2k_3)\Gamma(1+k_2k_3+k_1k_3)
\Gamma(1+k_1k_2+k_1k_3)}\ ,
\label{pole}
\ee
where the on-shell condition is $1+k_1k_2+k_2k_3+k_1k_3=0$.
By means of the on-shell condition, from the above expression,
we recover, up to a normalization constant, 
the Veneziano amplitude, the scattering amplitude of four on-shell
tachyons. Eq.(\ref{pole}) in fact becomes 
\bea
&&\frac{1}{\pi^3}
\Gamma(1+2k_1k_2)\Gamma(1+2k_2k_3)\Gamma(1+2k_1k_3)
\sin(\pi k_1)\sin(\pi k_2)\sin(\pi k_3)\cr
&&=\frac{1}{(2\pi)^2}\left[B\left(1+2k_1k_2,1+2k_2k_3\right) 
+{\rm cycl.}\right]\ ,
\label{ven}
\eea 
where $B(x,y)$ is the Euler beta function.
The expression between square brackets 
is just the Veneziano amplitude.
The ambiguity $c$ appearing in the propagator (\ref{prop0}) could be kept undetermined
throughout the calculations of the scattering amplitudes.
It is not difficult to see that this would just consistently change the normalization
of the on-shell amplitudes.

For tachyon profiles $T_R(k)$
supported over near on-shell momentum $k^2\simeq 1$, the equation of motion $\beta^T=0$
with $\beta^T$ given in (\ref{betafunction}) becomes
\bea
\beta^T(k)&=&(1-k^2)T_R(k)-\frac{(2\pi)^D}{2\pi}\int dk_1 dk_2
 \delta (k-k_1-k_2)T_R(k_1) T_R(k_2)
\cr
&+&\frac{(2\pi)^D}{3!(2\pi)^2}\int dk_1 dk_2 dk_3\delta(k-k_1-k_2-k_3)
T_R(k_1) T_R(k_2) T_R(k_3)\cr
&&\left\{\left[B\left(1+2k_1k_2,1+2k_2k_3\right) 
+{\rm cycl.}\right]+2\pi\tan(\pi k_1k_2)\tan(\pi k_1k_3)\tan(\pi k_2k_3)\right\}=0\ .\cr
&&
\label{betaos}
\eea
The coefficients of the quadratic and cubic terms in (\ref{betaos})
are symmetric with respect to all the $k_i$ and $k$ when
these are on the mass-shell. Thus (\ref{betaos}) can be derived as
the equation of motion of an effective action.
Such effective action for near on-shell tachyons up to the
fourth order in powers of the tachyon fields can be derived from the results
of the cubic string field theory. In~\cite{Taylor:2002bq}
it was shown that the cubic SFT reproduces the Veneziano amplitude 
with great accuracy already at level $L=50$. The tachyon effective action 
arising from the cubic string field theory for near on shell tachyon profiles $\Phi(k)$
therefore reads
\bea
S_C&&=2\pi^2T_{25}(2\pi)^{D}\left\{-\frac{1}{2}\int dk \Phi(k) \Phi(-k) \left(1-k^2\right)
+\frac{1}{3}\int \prod_{i=1}^3dk_i \Phi(k_i)\delta\left(\sum_{i=1}^3 k_i\right)
\right.\cr+
&&\left.\frac{1}{4!}\int \prod_{i=1}^4 dk_i \Phi(k_i)\delta\left(\sum_{i=1}^4 k_i\right)
\left[B\left(1+2k_1k_2,1+2k_2k_3\right) + {\rm cycl.}\right]\right\}\ ,\cr
&&
\label{cubicos}
\eea
where the tachyon momenta in the fourth order term satisfy
\begin{eqnarray}
k_1 = (0, 1, 0, 0, \ldots, 0) & \hspace{1in} & 
k_2 = (0, \sin \theta,  \cos \theta, 0, \ldots, 0) \\
k_3 = (0, -1, 0, 0, \ldots, 0) & \hspace{1in} & 
k_4 = (0, -\sin \theta,  -\cos \theta, 0, \ldots, 0)\ .
\end{eqnarray}
Since the Veneziano amplitude is completely symmetric in the four momenta $k_i$, it is not
difficult to see that the equation of motion deriving from (\ref{cubicos})
becomes precisely (\ref{betaos}) once the simple field rescaling $T=2\pi \Phi$ is 
performed. Thus the cubic string field theory for almost on-shell tachyons
reproduces the non-linear $\beta^T=0$ equation of motion.

In section 4 we also 
derived the renormalized tachyon field for the case of a slowly varying 
tachyon profile, to all orders in the bare field and to the leading order
in derivatives, eq.(\ref{trsum}). From this we can easily compute the corresponding $\beta$
function. The task in this case is much simpler, as we just need to take the 
derivative of (\ref{trsum}) with respect to $-\log\epsilon$
\be
\beta(X)=\frac{\partial T_R(X)}{\partial(-\log\epsilon)}=
\frac{1}{\epsilon}\exp\left(-\frac{T(X)}{\epsilon}\right)\left\{T(X)+
\triangle T(X)\left[1-\left(1-\frac{T(X)}{\epsilon}\right)
\log\epsilon\right]\right\}\ .
\label{betax}
\ee
Then we have to invert the relation (\ref{trsum}) between $T_R$ and $T$.
To the leading order in derivatives one has
\be
T(X)=-\epsilon\left\{[1+(\log\epsilon)\triangle]\log(1-T_R(X))\right\}\ ,
\label{ttr}
\ee
from which it is clear that the admissible range for $T_R$ is 
$-\infty\le T_R\le 1 $. Plugging (\ref{ttr}) into (\ref{betax})
we get
the non-linear tachyon $\beta$-function to all powers 
of the renormalized tachyon and to the leading order in its derivatives 
\be
\beta^T(X)=\left(1-T_R(X)\right)\left[-\log\left(1-T_R(X)\right)-
\triangle\log\left(1-T_R(X)\right)\right]\ .
\label{betaexact}
\ee
$\beta^T(X)=0$ is  the tachyon equation of motion
for a slowly varying tachyon profile.

Since in our calculations of the non-linear $\beta$-function we have always used the same coordinates
in the space of string fields, the two results (\ref{betaexact}) and (\ref{betafunction})
should coincide when expanded up to the third power of the field and 
to the leading order in derivatives, respectively. This is indeed the case
and the result in both cases reads
\be
\beta^T(X)=\triangle T_R+\partial_\mu T_R\partial_\mu 
T_R+T_R\partial_\mu T_R\partial_\mu T_R\ .
\ee

It is interesting to compute the $\beta$-function also in the case in which the
ambiguity constant $c$ appearing in (\ref{prop0}) is kept undetermined.
$T_R(k)$ can be easily obtained as before from
(\ref{zetanexp}) without fixing $c=4$. By taking the Fourier transform
and by differentiating with respect to $-\log\epsilon$, the $\beta$-function
expressed in terms of the renormalized tachyon field $T_R(X)$ turns out to be
\be
\beta^T(X)=\left(1-T_R\right)\left[-\log\left(1-T_R\right)
+\frac{\triangle T_R}{1-T_R}+\left(1+\frac{1}{2}
\log\frac{c}{4}\right)\frac{\partial_\mu T_R\partial_\mu T_R}{(1-T_R)^2}\right]\ .
\label{betaexactc}
\ee
In the next section we shall use also this form of the $\beta$-function
to construct the Witten-Shatashvili action.

\section{Witten-Shatashvili action}

In this section we shall compute the Witten-Shatashvili action.
{}From the simple expression that relates the partition function 
to the renormalized tachyon (\ref{zt}) it is easy to deduce a 
simple and universal form for the WS action of the open bosonic string 
theory
\begin{equation}
S=\left( 1-\int \beta^T\frac{\delta}{\delta T_R}\right)Z[T_R]
=K\int d^DX\left[1-T_R(X)+\beta^T(X)\right]\ .
\label{witten1}
\end{equation}
This can now be computed in both the cases analyzed in the previous sections.
We shall show that the expressions for $S$ that we will obtain 
are consistent both with the known results on the tachyon potential~\cite{Kutasov:2000qp}
and with the expected on-shell behavior. Thus a choice of coordinates in
the space of couplings in which the tachyon $\beta$-function is 
non-linear allows one to find not only a simple general formula for the
WS action, but provides also a space-time tachyon effective action that
describes tachyon physics from the far-off shell to the near on-shell regions. 

Let us start with the evaluation of (\ref{witten1})
up to the third order in the expansion of the tachyon field
using the non-linear $\beta$-function (\ref{betafunction}).
A similar computation was done in~\cite{Kutasov:2000qp,Frolov:2001nb}
by means of the linear $\beta$-function, $\beta(k)= \left(1-k^2\right)T(k)$.
We shall later compare the two results.
{}From the renormalized field (\ref{tr3}) and the  $\beta$-function
(\ref{betafunction}) we arrive at the following expression for the 
Witten action
\bea
&&S=K\left\{
1-\frac{1}{2}\int dk(2\pi)^D T_R(k) T_R(-k) 
\frac{\Gamma(2-2k^2)}{\Gamma^2(1-k^2)}\right.\cr
&&\left.+
\frac{1}{3!}\int dk_1 dk_2 dk_3(2\pi)^{D}
T_R(k_1) T_R(k_2) T_R(k_3)\delta(k_1+k_2+k_3)\right.\cr
&&\left.\left[2\left(1+\sum_{i<j}k_ik_j\right)I(k_1,k_2,k_3)-\left(\frac{\Gamma(1+2k_2k_3)
\Gamma(2+2k_1k_2+2k_1k_3)}
{\Gamma^2(1+k_2k_3)\Gamma^2(1+k_1k_2+k_1k_3)}+{\rm cycl.}
\right)\right]\right\} .\cr
&&
\label{wittenaction}
\eea
The propagator coming from the quadratic term in (\ref{wittenaction})
exhibits the required pole at $k^2=1$. There are however also an
infinite number of other zeroes and poles.  
We shall show that these are due to the metric in the coupling space
appearing in (\ref{metric}).
The equations of motion derived from the action (\ref{wittenaction}) are
\bea
&&\frac{\delta S}{\delta T_R(-k)}=-K\frac{\Gamma(2-2k^2)}{\Gamma^2(1-k^2)}
(2\pi)^DT(k)\cr
&&+\frac{K}{2}\int dk_1dk' (2\pi)^D\delta\left(k_1+k'-k\right)T_R(k_1)T_R(k')\cdot \cr
&&\cdot \left\{
2(1-k_1k+k_1k'-kk')I(-k,k_1,k')-\frac{\Gamma(1-2kk_1)\Gamma(2-2kk'+2k_1k')}
{\Gamma^2(1-kk_1)\Gamma^2(1-kk'+k'k_1)}~\right.\cr
&&~~~\left.-\frac{\Gamma(1-2kk')\Gamma(2-2kk_1+2k'k_1)}
{\Gamma^2(1-kk')\Gamma^2(1-kk_1+k'k_1)}-\frac{\Gamma(1+2k'k_1)\Gamma(2-2kk'-2kk_1)}
{\Gamma^2(1+k'k_1)\Gamma^2(1-kk'-kk_1)}\right\}\ .
\label{eom}
\eea
As we did for the equation $\beta^T=0$ in the previous section, by solving 
these equations perturbatively
it is possible to recover the scattering amplitudes for three on-shell tachyons.
To the lowest order the equation is 
\be
\frac{\Gamma(2-2k^2)}{\Gamma^2(1-k^2)}T_0(k)=0\ .
\label{lowest}
\ee
At variance with the lowest order solution of $\beta^T=0$, there are infinite
possible solutions of (\ref{lowest}). We choose
the solution for which the tachyon field $T_0(k)$ is on the
mass-shell, which corresponds to a consistent string theory background.
This choice is also a solution  of $\beta^T=0$ to the lowest order.
As we shall show, the other possible zeroes of (\ref{lowest})
could be interpreted as zeroes of the metric in the space of couplings
through eq.(\ref{metric}).
With such a choice of $T_0(k)$, to the next order we recover 
the scattering amplitudes for three on-shell tachyons. 
By writing $T(k)=T_0(k)+T_1(k)$ and substituting it into 
(\ref{eom}) we find
\bea
&&T_1(k)=\frac{\Gamma^2(1-k^2)}{2\Gamma(2-2k^2)}\int dk_1 dk'
(2\pi)^D\delta(k-k_1-k')T_0(k_1)T_0(k')
\cr
&&\left\{
2(1-k_1k+k_1k'-kk')I(-k,k_1,k')-\frac{\Gamma(1-2kk_1)\Gamma(2-2kk'+2k_1k')}
{\Gamma^2(1-kk_1)\Gamma^2(1-kk'+k'k_1)}~\right.\cr
&&~~~\left.-\frac{\Gamma(1-2kk')\Gamma(2-2kk_1+2k'k_1)}
{\Gamma^2(1-kk')\Gamma^2(1-kk_1+k'k_1)}-\frac{\Gamma(1+2k'k_1)\Gamma(2-2kk'-2kk_1)}
{\Gamma^2(1+k'k_1)\Gamma^2(1-kk'-kk_1)}\right\}\ .
\label{next}
\eea
Since the two couplings $T_0$ satisfy the on-shell condition, $k_1$ and $k'$ are
on-shell. To pick out the propagator pole corresponding to the
third $k$ we set it on-shell too.
The scattering amplitude for three on-shell tachyons is given again by
the residue of the pole and  with our normalization is $(2\pi)^{-1}$, in precise agreement with
the result obtained in the previous section. 

The equations (\ref{eom})
must be related to the equation $\beta^T=0$ through a metric 
$G_{T(k)T(k')}$ as in (\ref{metric}), which in this case becomes
\be
\frac{\delta S}{\delta{T_R(k)}}=-\int dk' G_{T(k)T(k')}\beta^{T(k')}\ .
\label{ds}
\ee
The Witten-Shatashvili formulation of string field theory provides a prescription
for the metric $G_{T(k)T(k')}$ which can then be computed explicitly.
To the first two orders in powers of $T_R$, it is given by
\bea
G_{T(k)T(k')}&&=K\frac{(2\pi)^D\Gamma(2-2k^2)}{(1-k^2)\Gamma^2(1-k^2)}\delta\left(k+k'\right)
-\frac{K}{2}\int dk_1 (2\pi)^D\delta\left(k+k'+k_1\right)\frac{T_R(k_1)}{1-k'^2}\cdot \cr
&&\cdot \left\{
2(1+k_1k+k_1k'+kk')I(k_1,k,k')-\frac{\Gamma(1+2kk_1)\Gamma(2+2kk'+2k_1k')}
{\Gamma^2(1+kk_1)\Gamma^2(1+kk'+k'k_1)}~\right.\cr
&&~~~\left.-\frac{\Gamma(1+2kk')\Gamma(2+2kk_1+2k'k_1)}
{\Gamma^2(1+kk')\Gamma^2(1+kk_1+k'k_1)}-\frac{\Gamma(1+2k'k_1)\Gamma(2+2kk'+2kk_1)}
{\Gamma^2(1+k'k_1)\Gamma^2(1+kk'+kk_1)}  ~\right.\cr
&&~~~\left.-\frac{\Gamma(2+2k'k_1)\Gamma(2+2kk'+2kk_1)}
{\Gamma^2(1+k'k_1)\Gamma^2(1+kk'+kk_1)(1+kk'+kk_1)}\right\}\ .
\label{gmetric}
\eea
The first term in this metric coincides with (\ref{gij})
for a conformal primary given by the tachyon vertex.
{}From (\ref{gmetric}) it is possible to see that the infinite
number of zeroes and poles that the second order term in eq.(\ref{wittenaction}) exhibits
at $k^2=1+n$ and $k^2=3/2+n$, respectively, is in fact due to the metric.
This is true except for the zero corresponding to the tachyon mass-shell $k^2=1$. 
In fact the metric (\ref{gmetric}) is regular for $k^2=1$. This indicates that the
kinetic term in eq.(\ref{wittenaction}) exhibits the required zero at the tachyon mass-shell 
and the metric (\ref{gmetric}) can be made responsible for the other extra zeroes and poles.
If these zeroes and poles are just an artifact of the expansion in powers of $T$,
it is an open question. It would be interesting to consider for example an expansion 
around $k^2=1+n$ to all orders in $T$ and check if in this case one would still find that
the kinetic term exhibits a zero at $k^2=1+n$.

Let us turn now to the cubic term in eq.(\ref{wittenaction}).
If one or two tachyons are on-shell, then the cubic term vanishes. 
This means that any exchange diagram 
involving the cubic term vanishes~\cite{Frolov:2001nb}.
When all the three tachyons are on-shell, 
the scattering amplitude for three on-shell tachyons should arise directly
as the coefficient of the cubic term. However,
the cubic term in (\ref{wittenaction})  is ill-defined on shell.
Nonetheless, with the
most obvious regularization ({\it i.e.}  by going on-shell
symmetrically by giving to the three tachyons an identical 
small mass $m$,  $k_i^2=1+m^2$ and then by taking the $m\to 0$
limit) one gets a finite result for the scattering amplitude~\cite{Frolov:2001nb}.
Recalling the first of eqs.(\ref{coeff}) we conclude that 
this scattering amplitude is $(2\pi)^{-1}$ with our normalization.
Also the cubic term in (\ref{wittenaction}) has a sequence of poles 
at finite distances from the tachyon mass-shell. This is related 
to the fact that the set of couplings that we have taken into account is 
not complete. If we get far enough from the tachyon mass-shell, 
we run into the poles due to all the other string states which 
have not been subtracted.
 
In the next section 
we shall compare (\ref{wittenaction}) with 
the corresponding action derived from the cubic string field theory.
Here we would like to show that, by means of a field 
redefinition, (\ref{wittenaction}) can be rewritten in the form of the WS action obtained from
a linear $\beta$-function~\cite{Frolov:2001nb}, 
but that this field redefinition becomes singular on-shell.
The partition function up to the third order in the bare tachyon field is again given
by
\bea
&&Z(k)=K\delta(k)-K\epsilon^{k^2 -1}\left[T(k)\right.\cr
&&\left.-\frac{1}{2}
\int dk_1dk_2(2\pi)^{D}\delta \left(k-k_1-k_2\right)\epsilon^{-(1+2k_1k_2)} 
{T}(k_1){T}(k_2)\frac{\Gamma \left(1+2k_1k_2\right)}
{\Gamma^2 \left(1+k_1k_2\right)} \right. \cr
&&\left. +\frac{1}{3!}
\int dk_1dk_2dk_3(2\pi)^{D}\delta \left(k-\sum_{i=1}^3k_i\right)
\epsilon^{-2(1+\sum_{i<j}k_ik_j)}
{T}(k_1){T}(k_2)T(k_3)
I(k_1,k_2,k_3)\right]\ ,\cr
&&
\label{z(k)}
\eea
where we have used (\ref{tr3}). If instead of following the general procedure of
ref.~\cite{Klebanov:1987wx} one renormalizes the theory
simply by normal ordering, the $\beta$-function turns out to be linear. 
Thus the renormalized field to all orders
in the bare field would just be
\be
\phi_R(k)=T(k)\epsilon^{k^2-1}\ ,
\ee
so that  $\beta(k)= \left(1-k^2\right)\phi_R(k)$.
The WS action with a linear $\beta$-function up to the third order in the 
tachyon field then reads
\bea
&&S_L=K\left\{
1-\frac{1}{2}\int dk(2\pi)^D \phi_R(k) \phi_R(-k) 
\frac{\Gamma(2-2k^2)}{\Gamma^2(1-k^2)}\right.\cr
&&\left.+
\frac{1}{3!}\int dk_1 dk_2 dk_3(2\pi)^{D} 
\phi_R(k_1) \phi_R(k_2) \phi_R(k_3)\delta\left(\sum_{i=1}^3k_i\right)
2\left(1+\sum^3_{i<j=2}k_ik_j\right)I(k_1,k_2,k_3)\right\}\cr
&&
\label{wittenaction2}
\eea
in agreement with what found in~\cite{Frolov:2001nb}.
If we assume that the fields $\phi_R$ and $T_R$ are related as
follows
\be
\phi_R(k)=T_R(k)+\int dk_1 f(k,k_1)T_R(k_1) T_R(k-k_1)+\dots\ ,
\ee
by comparing the cubic terms in (\ref{wittenaction}) and (\ref{wittenaction2})
one finds
\bea
&&\left[f(k_2+k_3,k_2)\frac{\Gamma(2+2k_1k_2+2k_1k_3)}
{\Gamma^2(1+k_1k_2+k_1k_3)}+{\rm cycl.}\right]\cr
&=&\frac{1}{2}\left[\frac{\Gamma(1+2k_2k_3)}{\Gamma^2(1+k_2k_3)}
\frac{\Gamma(2+2k_1k_2+2k_1k_3)}{\Gamma^2(1+k_1k_2+k_1k_3)}+{\rm cycl.}\right]\ ,
\eea
so that the solution for $f$ is $f(k_1+k_2,k_1)={\Gamma(1+2k_1k_2)}/(2 \Gamma^2(1+k_1k_2))$ and the
field redefinition becomes
\be
\phi_R(k)=T_R(k)+\int dk_1 dk_2 \frac{\Gamma(1+2k_1k_2)}{2 
\Gamma^2(1+k_1k_2)}T_R(k_1) T_R(k_2)\delta(k-k_1-k_2)\ .
\label{phirtr}
\ee
It is not difficult to see that if we evaluate this relation when
the three tachyon fields are on-shell it becomes singular since 
$f(k,k_1)$ has a pole.
 This is in agreement with the 
Poincar\'e-Dulac theorem~\cite{Zamolodchikov:1989zs} 
used in~\cite{Shatashvili:1993ps} to prove that when the resonant condition (\ref{resonant})
holds, namely near the on-shellness, the $\beta$-function has to be non-linear.
We showed in fact that the field redefinition that gives from $S$ the WS action
constructed in terms of a linear $\beta$-function, $S_L$,
becomes singular on-shell.

Let us now turn to the WS action computed in an expansion to the leading order in derivatives
and to all orders in the powers of the tachyon fields. 
If we  keep the renormalization ambiguity $c$ undetermined,
the $\beta$-function is  
given in (\ref{betaexactc}). Using (\ref{sws}), $S$ then reads
\be
S=K\int dX\left(1-T_R\right)\left[1-\log\left(1-T_R\right)
+\left(1+\frac{1}{2}
\log\frac{c}{4}\right)\frac{\partial_\mu T_R\partial_\mu T_R}{(1-T_R)^2}\right]\ ,
\label{wssv}
\ee
where $-\infty\le T_R\le 1$. With the field redefinition
\be
1-T_R=e^{-\tilde T}
\label{fr}
\ee
$S$ becomes
\be
S=K\int dXe^{-\tilde T}\left[\left(1+\frac{1}{2}
\log\frac{c}{4}\right)\partial_\mu\tilde T\partial_\mu\tilde T+1+\tilde T\right]\ ,
\label{wssvr}
\ee
which for $c=4$ coincides with the space-time tachyon action found in~\cite{Gerasimov:2000zp,Kutasov:2000qp}.
In particular we shall show in the next section that $K$ coincides with the tension of the 
D25-brane, $K=T_{25}$, 
in agreement with the results of ref.~\cite{Kutasov:2000qp}.
It is not difficult to show that (\ref{wssvr}) can be rewritten, by
means of a field redefinition, in the form found in~\cite{Tseytlin:2000mt}
where the renormalization ambiguity was also discussed.

Note that (\ref{fr}) is the coordinate transformation in the coupling space
that leads form the non-linear $\beta$-function
(\ref{betaexact}) to the linear beta function $\beta^T=(1+\triangle)T$.
The $\beta$-function in fact is a covariant vector in the coupling space
and as such it transforms. 

We have left the ambiguity $c$ in (\ref{wssvr}) undetermined because we want
to show that it is possible to fix $c$ in such a way that the equation of 
motion deriving from (\ref{wssvr}) coincides with the equation $\beta^T=0$
with $\beta^T$ given in (\ref{betaexactc}). In fact, 
 in terms
of the coordinates (\ref{fr}), this equation reads 
\be
\beta^{\tilde T}=\tilde T+\triangle \tilde T+\frac{1}{2}
\log\frac{c}{4}\partial_\mu \tilde T\partial_\mu \tilde T =0\ .
\label{betac}
\ee
where we have kept into account that $\beta^{\tilde T}$ transforms 
like a covariant vector in the space of worldsheet theories.
Choosing $\log({c}/{4})=-1$, eq.(\ref{betac}) becomes the
equation of motion of the action 
(\ref{wssvr}).
This is important because if we find finite action solutions of the equation
(\ref{betac}), these would be at the same time solutions of the renormalization
group equations $and$ solitons of the tachyon effective action (\ref{wssvr}). 
These could then be interpreted as lower dimensional branes. Being solutions 
of the renormalization group equations they are interpreted as background
consistent with the string dynamics, being solitons they must describe branes.
The finite action solutions of eq.(\ref{betac}) are easy to find 
\be
\tilde T(X)=-n+\frac{1}{2}\sum_{i=1}^n(X^i)^2\ .
\label{profile}
\ee
These codimension $n$ solitons
can be interpreted as D($25-n$)-branes. $26-n$ are in fact
 the number of coordinates on which the profile $\tilde T(X)$ does not depend. 
Substituting the solution (\ref{profile}) into the action (\ref{wssvr})
with $\log({c}/{4})=-1$ we get
\be
S=T_{25}(e\sqrt{2\pi})^n V_{26-n}\ .
\ee
Comparing this with the expected result $T_{25-n} V_{26-n}$ we derive the following 
ratio between the brane tensions
\be
R_n=\frac{T_{25-n}}{T_{25}}=(\frac{e}{\sqrt{2\pi}} 2\pi)^n\ .
\label{tension}
\ee
With our notation, $\alpha'=1$, the exact tension ratio should be $R_n=(2\pi)^n$.
Thus $R_n$ differs from the one given in (\ref{tension})
by a factor $e/\sqrt{2\pi}=1.084$. It is remarkable that a small derivatives
expansion of the WS action truncated just to the second order provides a 
result with the 93$\%$ of accuracy. In particular the result (\ref{tension})
is much closer to the exact tension ratio then the one found in~\cite{Kutasov:2000qp} 
with analogous procedure. 
The solutions of the equations of motion of the WS action considered in~\cite{Kutasov:2000qp} 
were not in fact solutions of the equation $\beta^T=0$, so that they could not
be interpreted as consistent string backgrounds (this was already noticed by the authors
of~\cite{Kutasov:2000qp} and for this reason the exact tension ratio was obtained with a 
different procedure).
The equations of motion deriving from
the WS action are in fact related to the $\beta$-function through (\ref{metric})
where the metric should in principle be non-degenerate. However, if the metric is
computed in some approximation, it could be singular and  present solutions that
introduce physics beyond that contained in the $\beta$-functions. 
The action (\ref{wssvr}) with $\log({c}/{4})=-1$ gives an equation of the 
form (\ref{metric}) with the non-degenerate metric $e^{-\tilde T}$. The solution of
this equation can be at the same 
time a soliton and a conformal RG fixed point.

In conclusion the general formula (\ref{sws}) reproduces all the expected results
on tachyon effective actions both in the far off-shell and in the near on-shell
regions.

\section{Cubic vs. Witten-Shatashvili tachyon effective actions}

In this section we shall compare the result (\ref{wittenaction}),
which gives the WS action up to the third order in the powers 
of the renormalized tachyon field $T_R(k)$, with the tachyon effective action
$S_C$ computed with the cubic open string field theory of~\cite{Witten:1985cc}.
Like (\ref{wittenaction}), $S_C$ is known exactly up to the third power 
in the tachyon field.
A similar comparison was already done in~\cite{Kutasov:2000qp} where,
however, the WS action constructed in terms of the linear tachyon 
$\beta$-function $\beta^T(k)=(1-k^2)T_R(k)$ was used. 
With such a choice of coordinates in the space of string fields, 
the relation between the tachyon fields of the cubic and the WS string 
field theory becomes singular on-shell~\cite{Kutasov:2000qp}.
The cubic string field theory parametrization of worldsheet RG is regular
on-shell and it very well reproduces the tachyon scattering amplitudes~\cite{Taylor:2002fy}, 
thus indicating that to it should correspond a non-linear $\beta$-function.
We shall show that, comparing the result (\ref{wittenaction})
for the WS action with the corresponding cubic string field theory action,
a  field redefinition between the tachyon fields in the two formulations
can be found which is non-singular on-shell.
In particular, by requiring the regularity of the coordinate transformation
that links the cubic tachyon effective action to the WS action (\ref{wittenaction})
we find that the overall normalization constant $K$ in the WS action
(\ref{wittenaction}) is precisely the tension of the D25-brane. This is in agreement
with all the conjectures involving tachyon condensation and with the result
$K=T_{25}$ derived from the tachyon potential.

For a tachyon field $\Phi(k)$, the cubic string field theory action can be written as~\cite{Ohmori:2001am,Ghoshal:2000gt}
\bea
S_C&&=2\pi^2T_{25}\left[-\frac{1}{2}\int dk(2\pi)^D \Phi(k) \Phi(-k) \left(1-k^2\right)\right.\cr
&&\left.+\frac{1}{3}\int dk_1 dk_2 dk_3(2\pi)^{D}\delta(k_1+k_2+k_3)
\Phi(k_1) \Phi(k_2) \Phi(k_3)\left(\frac{3\sqrt{3}}{4}\right)^{3-k^2_1-k^2_2-k_3^2}\right]\ .\cr
&&
\label{cubicac}
\eea
The normalization factor $2\pi^2 T_{25}$ was derived in~\cite{Sen:1999xm}
and will be important for our analysis.
Let us assume that the relation between the fields $\Phi(k)$ and 
$T_R(k)$ of the two theories is of the form
\be
\Phi(k)=f_1(k)T_R(k)+\int dk_1dk_2f_2(k,k_1)T_R(k_1)T_R(k_2)\delta(k-k_1-k_2)+\cdots\ ,
\label{rel}
\ee
where $f_1(k)=f_1(-k)$ from the reality of the tachyon field.
The cubic string field theory action (\ref{cubicac}) becomes
\bea
S_C&&=2\pi^2T_{25}\left\{-\frac{1}{2}\int dk(2\pi)^D T_R(k) T_R(-k) \left(1-k^2\right)\left(f_1(k)\right)^2
\right.\cr
&&\left.-\int dk_1 dk_2 dk_3(2\pi)^{D}\delta(k_1+k_2+k_3)T_R(k_1) T_R(k_2) T_R(k_3)\right.\cr
&&\left.\left[\left(1+k_1k_2+k_1k_3\right)f_1(k_1)f_2(k_2+k_3,k_3)
-\frac{1}{3}
f_1(k_1)f_1(k_2)f_1(k_3)\left(\frac{3\sqrt{3}}{4}\right)^{3-\sum_i k_i^2}\right]\right\}.\cr
&&
\label{cubicact}
\eea
By comparing the second order term of eq.(\ref{cubicac}) with the corresponding
term of eq.(\ref{wittenaction}) we find
\be 
\left(f_1(k)\right)^2=\frac{K}{2\pi^2T_{25}}\frac{\Gamma(2-2k^2)}{(1-k^2)\Gamma^2(1-k^2)}\ .
\label{f1}
\ee
When the tachyon field is on the mass-shell, $f_1(k)$ is regular and takes the value
\be
f_1=\frac{1}{2\pi}\sqrt{\frac{K}{T_{25}}}\ .
\label{f1os}
\ee
{}From the comparison of the cubic terms in (\ref{cubicact}) and in (\ref{wittenaction})
we get
\bea
&&\left[1-(k_2+k_3)^2\right]f_2(k_2+k_3,k_3)=\frac{f_1(k_2)f_1(k_3)}{3}
\left(\frac{3\sqrt{3}}{4}\right)^{3-\sum_ik_i^2}
-\frac{K}
{3!2\pi^2T_{25}f_1(k_2+k_3)}\cr
&&\left[2(1-k_2k_3-k^2_2-k^2_3)I(-k_2-k_3,k_2,k_3)
-\left(\frac{\Gamma(1+2k_2k_3)\Gamma(2-2(k_2+k_3)^2)}
{\Gamma^2(1+k_2k_3)\Gamma^2(1-(k_2+k_3)^2)}+{\rm cycl.}\right)\right].\cr
&&
\label{f2}
\eea
We can fix the value of the normalization
constant $K$ by requiring the regularity of the function $f_2(k_2+k_3,k_3)$ 
when the three tachyons are on-shell.
The on-shell condition is
$$
2(k_1k_2+k_2k_3+k_1k_3)+3=0\ ,
$$ 
and the factor between square brackets in (\ref{f2}) is, on-shell, 
$(2\pi)^{-1}$. 
Consequently, requiring the regularity of the function $f_2$ when the three tachyons are on-shell, 
eq. (\ref{f2}) simply becomes
\be
K=T_{25}\ .
\label{K}
\ee
When (\ref{K}) is satisfied, the field redefinition that
links the boundary and the cubic string field theory tachyon effective actions 
is regular on-shell. 
This result shows that the tachyon dynamics described by the WS string field theory
reproduces the conjectured relations involving tachyon condensation.
With analogous procedure,
it is not difficult to show
that the coordinate transformation between the WS tachyon effective action
constructed in terms of the linear $\beta$-function (\ref{wittenaction2}), 
with $K=T_{25}$, and (\ref{cubicac}) is, as expected, singular on-shell.

\section{Conclusions}

In this paper we have derived some exact results for the non-linear
tachyon $\beta$-function of the open bosonic string theory.
We have shown its relevance 
in the construction of the Witten-Shatashvili
bosonic string field theory.
When a non-linear renormalization 
of the tachyon field is considered~\cite{Klebanov:1987wx},
the WS action in fact is simply given by (\ref{sws}).
This formula has a wide range of validity. It
can be applied to the case in which the tachyon profile
is a slowly varying function of the embedding coordinates of the string
to derive the exact  tachyon  potential and the first derivative terms
of the effective action. Eq. (\ref{sws}) holds also when the tachyon 
coupling $T(k)$ is 
small and has support near the mass-shell. 
For such tachyon profiles we showed that perturbative solutions of the 
equation $\beta^T=0$ provide the expected scattering amplitudes of on-shell tachyons.

The explicit form of the WS action constructed from the tachyon non-linear 
$\beta$-function is in precise agreement with all the conjectures involving
tachyon condensation. In particular its normalization can be fixed either
by studying the exact tachyon potential or by finding the
field redefinition that maps the WS action into the effective tachyon action
coming from the cubic string field theory. This field redefinition is
non-singular on-shell only if the normalization constant coincides with
the tension of the D25-brane.

The knowledge of the non-linear tachyon $\beta$-function is very important 
also for another reason.
The solutions of the equation $\beta^T=0$ give the conformal fixed points,
the backgrounds that
are consistent with the string dynamics. In the case of slowly varying 
tachyon profiles,
we showed that the equations of motion for the WS action can be made
identical to the RG fixed point equation  $\beta^T=0$. This can be done for a 
particular choice of the renormalization prescription ambiguity.
We have found soliton solutions of this equation
to which correspond a finite value of the WS action.
Being solutions of the RG equations these solitons are 
lower dimensional D-branes for which the
finite value of $S$ provides a very accurate estimate of the D-brane tension.

When other excited string modes are present, say modes of the
vector field $A_\mu$, the form of
the WS action should still be given by (\ref{sws}) where 
the renormalized tachyon field depends also on the other string
couplings. In particular it would be interesting to apply
our renormalization procedure to the other renormalizable case 
in which the boundary perturbation contains also a vector field~\cite{coletti}.
Whether our approach would help in treating also non-renormalizable interactions
it is yet not clear.

The decay of unstable systems 
of D-branes, pictured as a tachyon field rolling down a potential 
toward a stable minimum, can also be addressed in the context of the
boundary string field theory. It involves deforming the world sheet 
conformal field theory of the unstable D-brane by a conformal, 
time dependent tachyon profile. It is useful then to 
construct $\beta$-functions which are valid away from the RG fixed point
to demonstrate that the renormalization flow exists, to
draw the RG-trajectories and to understand where the endpoints 
of the RG flux are. 
Thus our approach should reveal important in studying the
physics around a conformal fixed points and in particular about
the time dependent solutions describing rolling tachyons.

\acknowledgments{E.C. and G.G. would like
to thank R. Jackiw for an helpful discussion and the Bruno Rossi MIT-INFN 
fellowship for financial support. E.C. thanks
I. Sigalov and W. Taylor for useful discussions.
G.G. acknowledges
the M.I.T. Center for Theoretical Physics for
hospitality during the preparation of this work.}

\newpage

\appendix{\section{Computation of $I(k_1,k_2,k_3)$ }}

\begin{figure}[h]\vspace{-1.5cm}
\begin{center}
\epsfig{file=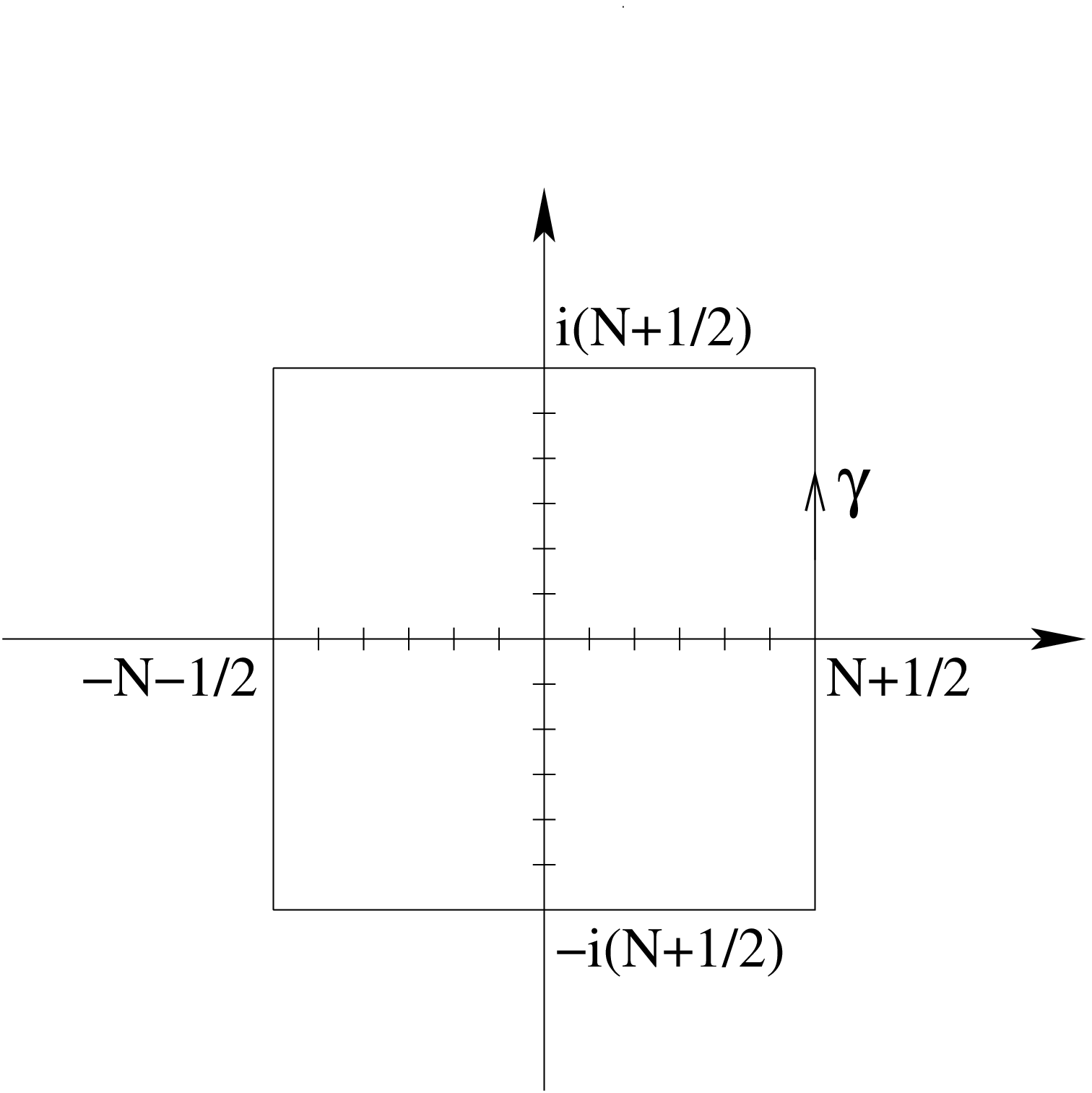,height=8cm}
\caption{Contour C.}
\label{figure}
\end{center}
\end{figure}

In this Appendix we shall compute the integral $I(k_1,k_2,k_3)$
appearing in eq.(\ref{z3})
\bea
I(k_1,k_2,k_3)=&&
\frac{2^{2k_1k_2+2k_2k_3+2k_1k_3}}{(2\pi)^3}
\int_0^{2\pi}d\tau_1d\tau_2d\tau_3\
\left[\sin^2\left(\frac{\tau_1 -\tau_2}{2}\right)
\right]^{k_1k_2}\cr
&&\left[\sin^2\left(\frac{\tau_2 -\tau_3}{2}\right)
\right]^{k_2k_3}\left[\sin^2\left(\frac{\tau_1 -\tau_3}{2}\right)
\right]^{k_1k_3}
\eea
Introducing the variables
$$
x=\frac{\tau_1 -\tau_2}{2}~~~,~~~y=\frac{\tau_3 -\tau_1}{2}
$$
the integral over $\tau_1$, $\tau_2$ and $\tau_3$
can be written as
$$
I=
-\frac{4^{k_1k_2+k_2k_3+k_1k_3+1}}{2\pi^3}\int_0^{2\pi}d\tau_1
\int_{\frac{\tau_1}{2}}^{\frac{\tau_1}{2}-\pi}dx
\int_{-\frac{\tau_1}{2}}^{\pi-\frac{\tau_1}{2}}dy
\left[\sin^2 x\right]^{k_1k_2}\left[\sin^2 y\right]^{k_1k_3}
\left[\sin^2(x+y)\right]^{k_2k_3}
$$
With a suitable shift of the integration variables
we obtain
\bea
I=&&
\frac{4^{k_1k_2+k_2k_3+k_1k_3}}{\pi ^2}
\int_0^{\pi}dx\int_0^{\pi}dy
\left[\sin x\right]^{2k_1k_2}\left[\sin y\right]^{2k_1k_3}
\left[\sin^2(x+y)\right]^{k_2k_3}\cr
=&&
\frac{4^{k_1k_2+k_2k_3}}{\pi^2}
\int_0^{\pi}dx\int_0^{\pi}dy\left[\sin x\right]^{2k_1k_2}
\left[\sin y\right]^{2k_1k_3}
\left[1-e^{2i(x+y)}\right]^{k_1k_3}\left[1-e^{-2i(x+y)}\right]^{k_2k_3}\cr
=&&
\frac{4^{k_1k_2+k_1k_3}}{\pi^2}
\int_0^{\pi}dx\int_0^{\pi}dy\left[\sin x\right]^{2k_1k_2}
\left[\sin y\right]^{2k_1k_3}
\sum_{n,m=0}^{\infty}
\frac{\Gamma(n-k_2k_3)\Gamma(m-k_2k_3)}{n!m!\Gamma^2(-k_2k_3)}
e^{2i(x+y)(n-m)}\cr
&&
\eea
Integrating over $x$ and $y$ we have
\bea
I=
\sum_{n,m=0}^{\infty}&&
\frac{\Gamma(n-a_3)\Gamma(m-a_3)}{n!m!\Gamma(1+a_1+n-m)
\Gamma(1+a_1-n+m)\Gamma(1+a_2+n-m)\Gamma(1+a_2-n+m)}\cr
&&\frac{\Gamma(1+2a_1)\Gamma(1+2a_2)}{\Gamma^2(-a_3)}
\eea
where $a_1=k_1k_2$, $a_2=k_2k_3$ and $a_3=k_1k_3$. 
Shifting $m\to m-n$ in the sum over $m$ we have
\bea
I&=&
\sum_{n,m=0}^{\infty}\frac{\Gamma(n-a_2)
\Gamma(n+m+a_2)\Gamma(1+2a_1)\Gamma(1+2a_3)}{n!(n+m)!\Gamma^2(-a_2)
\Gamma(1+a_1+m)
\Gamma(1+a_1-m)\Gamma(1+a_3+m)\Gamma(1+a_3-m)}\cr
&+&\sum_{n=0}^{\infty}\sum_{m=-n}^{0}\frac{\Gamma(n-a_2)
\Gamma(n+m+a_2)\Gamma(1+2a_1)\Gamma(1+2a_3)}{n!(n+m)!\Gamma^2(-a_2)
\Gamma(1+a_1+m)
\Gamma(1+a_1-m)\Gamma(1+a_3+m)\Gamma(1+a_3-m)}\cr
&-&\frac{\Gamma(1+2a_1)\Gamma(1+2a_3)}{\Gamma^2(1+a_1)\Gamma^2(1+a_3)}
~_2F_1(-a_2, -a_2;1;1)
\eea
where $~_2F_1\left(\alpha, \beta;\gamma;z\right)$
is the Hypergeometric function.
Changing the sign of the integer $m$ in the second term of the
previous equation and noting that
$$
\sum_{n=0}^{\infty}\sum_{m=0}^{n}=
\sum_{n=m}^{\infty}\sum_{m=0}^{\infty}
$$
we find
\bea
I&=&2
\sum_{m=0}^{\infty}\frac{\Gamma(m-a_2)\Gamma(1+2a_1)\Gamma(1+2a_3)}
{\Gamma(-a_2)\Gamma(1+a_1+m)
\Gamma(1+a_1-m)\Gamma(1+a_3+m)\Gamma(1+a_3-m)}\cr
&&~_2F_1(m-a_2, -a_2;m+1;1)
-\frac{\Gamma(1+2a_1)\Gamma(1+2a_3)}{\Gamma^2(1+a_1)\Gamma^2(1+a_3)}
~_2F_1(-a_2, -a_2;1;1)\cr
&&
\eea
It is not difficult to show that the sum over $m$ can be extended to
negative values so that we find
\bea
I=&&
-\frac{\Gamma(1+2a_1)\Gamma(1+2a_2)\Gamma(1+2a_3)}
{\Gamma(1-a_1)\Gamma(a_1)\Gamma(1-a_2)\Gamma(a_2)
\Gamma(1-a_3)\Gamma(a_3)}\cr
&&\sum_{m=-\infty}^{\infty}\frac{\Gamma(m-a_1)
\Gamma(m-a_2)\Gamma(m-a_3)}
{\Gamma(1+m+a_1)\Gamma(1+m+a_2)\Gamma(1+m+a_3)}
\label{i1}
\eea
The series in the second line of the right-hand side of (\ref{i1})
is convergent for $1+a_1+a_2+a_3>0$. 
To sum over $m$ we use a standard procedure. Consider the path in Fig.\ref{figure}
\vspace{-.5cm}\\
Defining 
\be
S=\sum_{m=-\infty}^{\infty}\frac{\Gamma(m-a_1)
\Gamma(m-a_2)\Gamma(m-a_3)}
{\Gamma(1+m+a_1)\Gamma(1+m+a_2)\Gamma(1+m+a_3)}\equiv\sum_{m=-\infty}^{\infty}f(m)
\ee
we can write
\be
\oint_C \pi {\rm cotg} \pi zf(z)dz=\sum_{m=-N}^{N}f(m)+\tilde S
\label{o}
\ee
where $\tilde S$ is the sum of the  residues of $ \pi {\rm cotg} \pi zf(z)$ 
evaluated in the poles of $f(z)$. 
In the limit $N \to \infty$ the left-hand side of the previous 
equation vanishes reducing $S$ to
\bea
S=&&
-\frac{\Gamma(1+a_1+a_2+a_3)}
{\Gamma(1+a_1)\Gamma(1+a_2)\Gamma(1+a_3)
\Gamma(1+a_1+a_2)\Gamma(1+a_1+a_3)\Gamma(1+a_2+a_3)}\cr
&&\left[\frac{\pi^3 \cos^2 \pi a_1}{\sin \pi a_1
\sin \pi (a_1-a_2)\sin \pi (a_1-a_3)}+{\rm cycl.} \right]
\label{s}
\eea
So that $I$ becomes
\be
I=
\frac{\Gamma(1+a_1+a_2+a_3)\Gamma(1+2a_1)\Gamma(1+2a_2)\Gamma(1+2a_3)}
{\Gamma(1+a_1)\Gamma(1+a_2)\Gamma(1+a_3)\Gamma(1+a_1+a_2)\Gamma(1+a_2+a_3)
\Gamma(1+a_1+a_3)}
\label{I}
\ee

\end{document}